%% file: paper.tex
\documentclass[sigplan,screen,prologue,table,usenames,dvipsnames,svgnames,x11names]{acmart}

\usepackage{epstopdf}
\usepackage{booktabs}
\usepackage[utf8]{inputenc}
\usepackage[T1]{fontenc}
\usepackage{xcolor}
\usepackage{pifont}
\usepackage{subcaption}
\usepackage[linesnumbered,ruled,vlined]{algorithm2e}

\SetCommentSty{mycommfont}
\usepackage{multirow}
\usepackage[abbreviations]{foreign}
\usepackage[]{hyperref} % always load last
\usepackage{tikz}
\usetikzlibrary{patterns,calc}
\usepackage{pgf-umlsd}
\usepackage{pifont}

\AtBeginDocument{
  \newcommand{\cmark}{\ding{51}}
  \newcommand{\xmark}{\ding{55}}
  \newcommand{\sys}{\textsc{PipeTune}\xspace}
  \newcommand{\baselinesys}{\textsc{Tune}\xspace}
  \definecolor{darkgreen}{rgb}{0.3,0.5,0.3}
  \newcommand{\yes}{{\color{darkgreen}\cmark}}
  \definecolor{darkred}{rgb}{0.5,0.3,0.3}
  \newcommand{\no}{{\color{darkred}\xmark}}
   
}

\usepackage{framed}
\colorlet{shadecolor}{Azure2}

\usepackage{cleveref} %ignored ?
\crefname{section}{§}{§§}

\usepackage[binary-units,per-mode=symbol]{siunitx}

\usepackage[inline]{enumitem}
\setlist{noitemsep,topsep=0pt,parsep=0pt,partopsep=0pt}
\usepackage{setspace}
\usepackage[margin=5pt,font={stretch=0.9}]{caption}

\begin{document}

\title[Pipeline Parallelism of Hyper and System Parameters Tuning for Deep Learning Clusters]{\sys: Pipeline Parallelism of Hyper and System Parameters Tuning for Deep Learning Clusters}

\author{Isabelly Rocha}
\affiliation{
  \institution{University of Neuchâtel}
  \city{Neuchâtel}
  \country{Switzerland}
}
\email{isabelly.rocha@unine.ch} 

\author{Nathaniel Morris}
\affiliation{
  \institution{The Ohio State University}
  \city{Columbus}
  \country{Ohio}
}
\email{morris.743@buckeyemail.osu.edu} 

\author{Lydia Y. Chen}
\affiliation{
  \institution{TU Delft}
  \city{Delft}
  \country{Netherlands}
}
\email{y.chen-10@tudelft.nl}

\author{Pascal Felber}
\affiliation{
  \institution{University of Neuchâtel}
  \city{Neuchâtel}
  \country{Switzerland}
}
\email{pascal.felber@unine.ch} 

\author{Robert Birke}
\affiliation{
  \institution{ABB Research}
  \city{Baden-D\"attwil}
  \country{Switzerland}
}
\email{robert.birke@ch.abb.com} 

\author{Valerio Schiavoni}
\affiliation{
  \institution{University of Neuchâtel}
  \city{Neuchâtel}
  \country{Switzerland}
}
\email{valerio.schiavoni@unine.ch} 

\renewcommand{\shortauthors}{I. Rocha, N. Morris, L. Y. Chen, P. Felber, R. Birke, V. Schiavoni}

\definecolor{darkgreen}{rgb}{0.3,0.5,0.3}
\definecolor{darkblue}{rgb}{0.3,0.3,0.5}
\definecolor{darkred}{rgb}{0.5,0.3,0.3}
%\definecolor{shadecolor}{rgb}{224,248,238}

%Comments by authors
\newboolean{showcomments}
\setboolean{showcomments}{true}
\ifthenelse{\boolean{showcomments}}{ 
  \newcommand{\mynote}[3]{
    \fbox{\bfseries\sffamily\scriptsize#1}
    {\small$\blacktriangleright$
     \textsf{\emph{\color{#3}{#2}}}$\blacktriangleleft$
     }
   }
}{\newcommand{\mynote}[3]{}}

\newcommand{\old}[1]{\mynote{Old}{#1}{red}}
\newcommand{\new}[1]{\mynote{New}{#1}{blue}}

\begin{abstract}
DNN learning jobs are common in today’s clusters due to the advances in AI driven services such as machine translation and image recognition. 
The most critical phase of these jobs for model performance and learning cost is the tuning of hyperparameters. 
Existing approaches make use of techniques such as early stopping criteria to reduce the tuning impact on learning cost. However, these strategies do not consider the impact that certain hyperparameters and systems parameters have on training time. 
This paper presents \sys, a framework for DNN learning jobs that addresses
the trade-offs between these two types of parameters.
PipeTune takes advantage of the high parallelism and
recurring characteristics of such jobs to minimize the learning cost via a pipelined simultaneous tuning of both hyper and system parameters. Our experimental evaluation using three different types of workloads indicates that \sys achieves up to 22.6\% reduction and $1.7\times$ speed up on tuning and training time, respectively. \sys not only improves performance but also lowers energy consumption up to 29\%.
\end{abstract}


\settopmatter{printacmref=false}
\setcopyright{none}
\renewcommand\footnotetextcopyrightpermission[1]{}
\pagestyle{plain}

\maketitle

\input{introduction}
\input{related}
\input{background}
\input{motivation}
%\input{problem}
\input{system}

\input{implementation}
\input{evaluation}
\input{conclusion}

\begin{acks}
We would like to thank our shepherd, Prof. Indranil Gupta, for his helpful feedback.
The research leading to these results has received funding from the European Union's Horizon 2020 research and innovation programme under the LEGaTO Project (\url{legato-project.eu}), grant agreement No~780681.
This research was partly funded by the SNSF NRP75 project Dapprox 407540\_167266.
\end{acks}

\bibliographystyle{ACM-Reference-Format}
\bibliography{biblio}

\end{document}

%% file: introduction.tex
%!TEX root = paper.tex

\section{Introduction}
\label{sec:introduction}

Deep Neural Networks (DNN) are becoming increasingly popular, both in academia and industry~\cite{kahng2017cti,dutta2018overview}.
They are being adopted across a variety of application domains, including speech~\cite{deng2013new,richardson2015deep, lozano2017analysis} and image recognition~\cite{erhan2014scalable}, self-driving vehicles~\cite{huval2015empirical}, face-recognition~\cite{strigl2010performance,sun2015deepid3}, genetic sequence modeling~\cite{xiong2015human}, natural language processing~\cite{deng2018deep}, e-health~\cite{10.1007/978-3-642-40763-5_51} and more.
Several public cloud providers offer native support to deploy, configure and run them, providing tools to automatically or semi-automatically drive the DNN processing pipeline. 
One important factor is the choice of the DNN hyperparameters (\eg, number of hidden layers, learning rate, dropout rate, momentum, batch size, weight-decay, epochs, pooling size, type of activation function,  \etc).
DNNs require careful tuning of the hyperparameters, and if done correctly, it can achieve impressive boosts in performance~\cite{sigopt-tensorflow-speedup, cvpr2018zoph}. 
However, misconfigurations can easily lead to wrong models and hence bad predictions~\cite{snow2018amazon,hao2019police}.

\begin{figure}[t]
    \centering
    \includegraphics[scale=0.65]{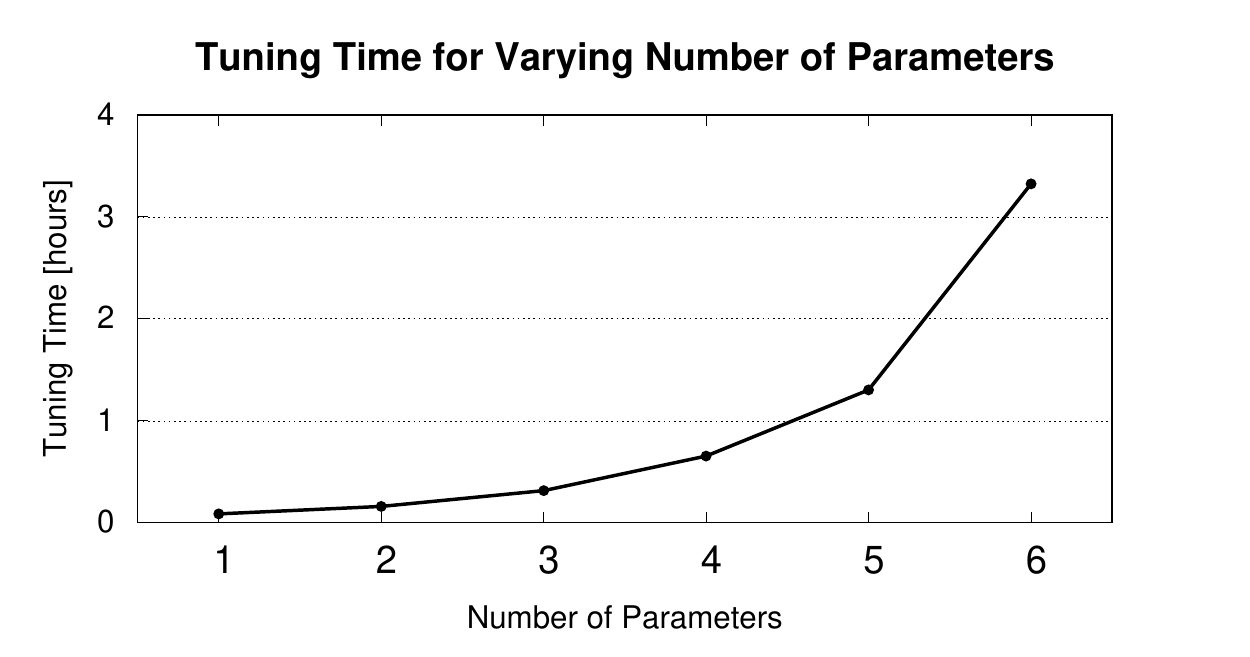}
    \includegraphics[scale=0.65]{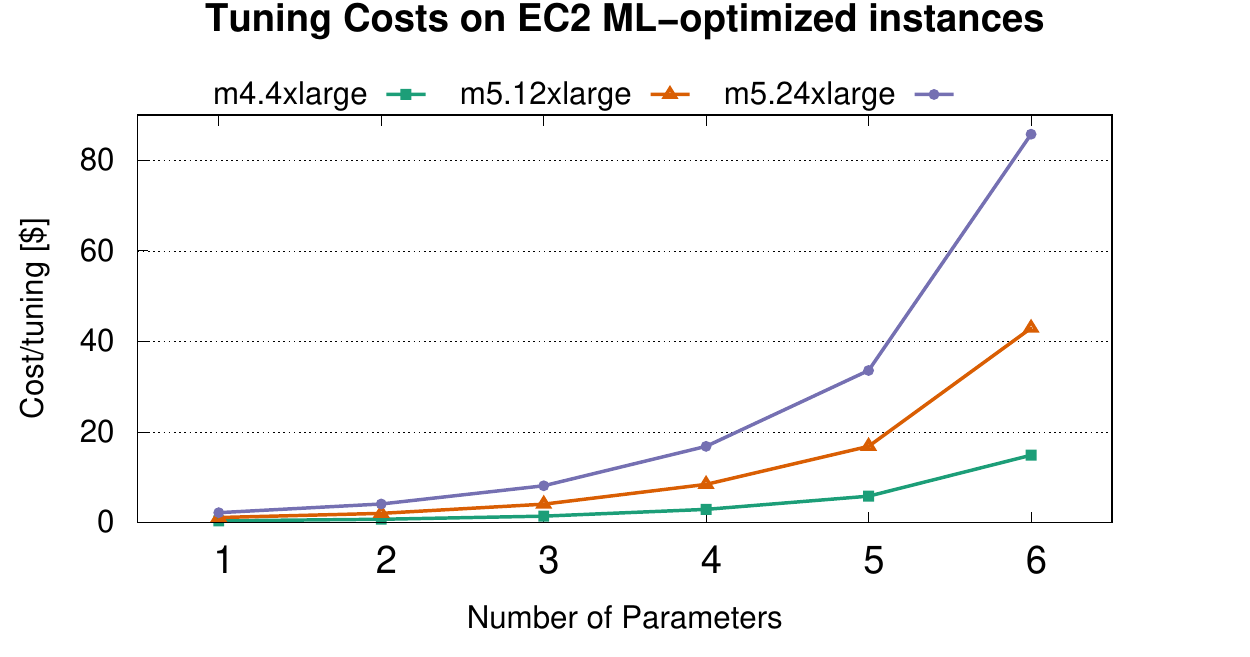}
    \caption{Clustering results grouped by workload type.}
    \label{fig:intro:tuningcosts}
\end{figure}

A naive approach to hyperparameter tuning is to perform a full exploration of the possible configuration variations.
Such a tuning approach becomes quickly unpractical, costly and slow, as the number of variations grows exponentially~\cite{pascanu2013difficulty}.
We show this using 3 types of ML-optimized EC2 instances in \Cref{fig:intro:tuningcosts} for a small number of parameters.
We take as example the tuning of a \textsc{LeNet} model on the \textsc{MSNIT} dataset and let it be tuned for different number of parameters (\ie, varying from 1 to 6). 
In this case, each parameter was configured to take up to 3 different values.
We measure the tuning time for each instance of this example and estimate the cost of doing so using a small, medium or large sized EC2 instance. 
We then observe that the cost of doing so grows exponentially with the number of parameters being tuned, becoming impractical.

Commercial platforms (\ie, Google Vizier~\cite{DBLP:conf/kdd/GolovinSMKKS17}, Amazon SageMaker~\cite{liberty2020elastic}), as well as on-premises solutions (\ie, Auto-Keras~\cite{jin2019auto}) help deployers by offering tuning services to mitigate (possibly avoid) misconfiguration. 

As a result of proper hyperparameters tuning, one should achieve fast convergence and high accuracy. 
Unfortunately, due to the tuning process length, this phase becomes expensive, and the situation exacerbates in cloud deployments~\cite{8814558}.
Even using cheap cloud instances (\ie, AWS EC2 Spot instances~\cite{10.1145/2509413.2509416}, as suggested for instance by AWS SageMaker~\cite{sagemaker}), the process can quickly lead to budget exhaustion.

We observe that some hyperparameters (\eg, number of epochs, batch size, dropout) can drastically reduce training time.
Importantly, training a DNN by using different system resources (\eg, number of CPU cores, allocated memory, number of GPUs) lead to different results, as we also demonstrate later in \Cref{fig:parameters} for varying number of cores.

However, handling system parameters as one of the hyperparameters is very time consuming, requiring in-depth knowledge of the workload, and it is often an intuition-driven process.
In addition, doing so would directly affect training and tuning time, and therefore state-of-the-art DNN tuning systems~\cite{bergstra2011algorithms} simply ignore this opportunity.
Instead, the majority of the existing tuning solutions restrict themselves to the sole hyperparameter tuning using a variety of techniques, including grid search~\cite{hinton2012practical}, random search~\cite{bergstra2012random}, hyperband~\cite{li2017hyperband}, bayesian optimization~\cite{shahriari2015taking,snoek2012practical}, evolutionary algorithms~\cite{10.1145/2834892.2834896, 10.1145/3071178.3071229}, population-based training (PBT)~\cite{jaderberg2017population}, \etc.
While a possible yet naive approach to treat system parameters is to consider them as possible hyperparameters, this leads to longer training periods (see \Cref{tab:compare}).

\begin{figure}[!t]
    \centering
    \includegraphics[scale=0.28]{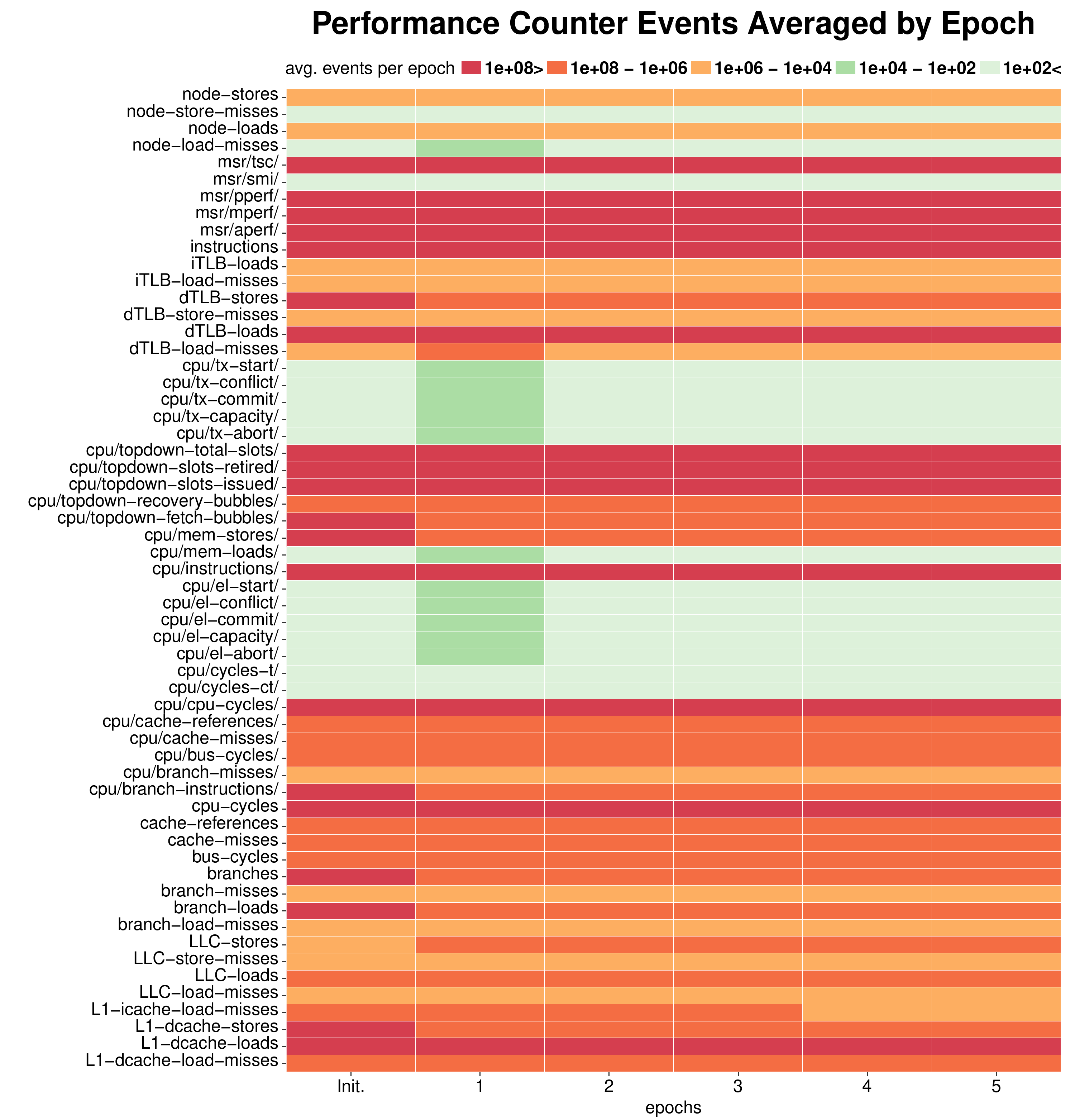}
    \caption{Profiling of training a \textsc{CNN} model on the \textsc{News20} dataset~\cite{news20} during the initiation phase and the 5 following epochs with 16 cores and 32GB memory.}
    \label{fig:probing}
\end{figure}

\sys strives to optimize both accuracy and training time of DNNs, while simultaneously tuning hyper and system parameters. 
The key observation of \sys is that the backbone of popular training algorithms for DNN is stochastic gradient decent~\cite{bengio2000gradient}, an iterative algorithm. 
\sys exploits such repetitive patterns as a unique opportunity to improve and achieve fast system parameter tuning. 
As an example, \Cref{fig:probing} illustrates the typical repetitive behavior of a training process.
We use a heatmap to show the hardware events happening through the training of a CNN model on the \textsc{News20} dataset~\cite{news20} during 5 epochs.
On the y-axis we show 58 different hardware events, on the y-axis initiation phase plus 5 training epochs. 
Each cell of the heatmap represents the average number of each event per single epoch.
We see how certain events repeat throughput the epochs with the same occurrence.

\input{rw_table}

Building on this observation, we design, implement and evaluate \sys, a middleware solution coordinating between the DNN training applications and systems.
In a nutshell, \sys relies on low level metrics to profile the training trials on the epoch level and make quick decisions regarding the system parameters.
The main research questions that \sys intends to answer, and the main contributions of this work are the following.

\begin{shaded*}
    \textit{\textbf{RQ1}: Why system parameters must be taken into account in the process of DNN tuning?}
\end{shaded*}
We show (\S~\ref{sec:background}) that by taking into account the system parameters, the overall tuning runtime can be greatly reduced while at the same time improving the model performance. 
Moreover, the training time can at the same time benefit from this approach, especially if the underlying system resources and their usage is exposed to the tuning phase.

\begin{shaded*}
    \textit{\textbf{RQ2}: Can out-of-the-box hyperparameter optimization algorithms also take care of system parameters?}
\end{shaded*}
We show that it is possible to include system parameters in the tuning process and ask the algorithm to optimize the ratio of accuracy to performance. 
However, our experimental evidences (\S~\ref{sec:evaluation}) highlight the following drawbacks. 
First, tuning runtime significantly increases (up to $\times$1.5 in our experiments).
Second, in doing so, the delicate equilibrium between performance and accuracy is negatively affected.

\textbf{Roadmap.} The reminder of this paper is organized as follows. 
We discuss related work and clarify how \sys positions in \S~\ref{sec:related}. 
In \S~\ref{sec:background}, we present a background of DNN tuning and outlines the basic features
needed to support \sys. 
In \S~\ref{sec:debunk}, we present an alternative approach relying on state-of-the-art solutions and show the need for our novel approach. 
We present the design of \sys in \S~\ref{sec:system} and describe its prototype implementation in \S~\ref{sec:implementation}. 
In \S~\ref{sec:evaluation}, we present the results of our in-depth evaluation.
Finally, we conclude in \S~\ref{sec:conclusion}.

%% file: rw_table.tex
%!TEX root = paper.tex

\begin{table*}[t]
  \caption{State-of-the-art systems related to hyper and system parameter tuning.}
  \label{tab:rw}
  \setlength{\tabcolsep}{3pt}
  \scriptsize 
  \rowcolors{1}{gray!10}{gray!0}
  \resizebox{0.95\textwidth}{!}{
  \begin{tabular}{l|c|c|c|c|cc|ccccc|c}
	 \rowcolor{gray!25}
    & & & & & \multicolumn{2}{l|}{\textbf{Parameter Tuning}} & \multicolumn{5}{c|}{\textbf{Supported DL Frameworks}} & \\ 
	\rowcolor{gray!25}
    \multirow{-2}{*}{\textbf{System}} & \multirow{-2}{*}{\textbf{CPU}} & \multirow{-2}{*}{\textbf{GPU}} & \multirow{-2}{*}{\textbf{Distributed}} & \multirow{-2}{*}{\textbf{Training}} & Hyper & System & BigDL & TensorFlow & Keras & PyTorch & MXNet & \multirow{-2}{*}{\textbf{Open Source}}\\ %Caffe? 
    \hline
    Astra~\cite{DBLP:conf/asplos/SivathanuCSZ19} & \no & \yes & \no & \yes & \yes  & \no & \no & \yes & \yes & \no & \no  & \no \\
    AutoKeras~\cite{jin2019auto} & \yes & \yes & \no & \yes & \yes  & \yes & \no & \yes & \yes & \no & \no  & \yes \\
    ByteScheduler~\cite{DBLP:conf/sosp/PengZCBYLWG19} & \yes & \yes & \yes & \yes & \no & \yes & \no & \yes & \yes & \yes & \yes  & \yes \\
    GRNN~\cite{DBLP:conf/eurosys/PengBCWG18} & \yes & \yes & \no  & \yes & \no & \no & \no & \yes & \no & \yes & \no  & \no \\
    HyperDrive~\cite{rasley2017hyperdrive} & \yes & \yes &  \yes & \yes & \yes & \no & \no & \yes & \yes & \no & \no & \no \\
    Hop~\cite{DBLP:conf/asplos/LuoLZQ19} & \yes & \no & \yes  & \yes & \no & \no & \no & \yes & \no & \no & \no  & \no \\
    Optimus~\cite{DBLP:conf/eurosys/PengBCWG18} & \yes & \yes & \yes  & \yes & \no & \no & \no & \no & \no & \no & \yes  & \no \\
    Orion~\cite{DBLP:conf/eurosys/WeiGGX19} & \yes & \no & \yes & \yes & \no  & \no & \no & \yes & \no & \no & \no  & \yes \\
    Parallax~\cite{DBLP:conf/eurosys/KimYPCJHLJC19} & \yes & \yes & \yes & \yes & \no  & \no & \no & \yes & \no & \no & \no  & \yes \\
    PipeDream~\cite{DBLP:conf/sosp/NarayananHPSDGG19} & \no & \yes & \yes & \yes & \no & \no & \no & \yes & \no & \no & \yes & \yes \\
    SageMaker~\cite{liberty2020elastic} & \yes & \yes & \yes & \yes & \yes & \no & \no & \no & \no & \no & \no  & \no \\
    STRADS~\cite{DBLP:conf/eurosys/KimHLZDGX16} & \yes & \no & \yes & \yes & \no & \no & \no & \no & \no & \no & \no & \yes  \\
    STRADS-AP~\cite{DBLP:conf/usenix/KimAGX19} & \yes & \no & \yes & \yes & \yes & \no & \no & \yes & \no & \no & \no  & \no \\
    Tune~\cite{liaw2018tune} & \yes & \yes & \yes & \yes & \yes & \no & \no & \yes & \yes & \no & \no  & \yes \\
    Vizier~\cite{DBLP:conf/kdd/GolovinSMKKS17} & \yes & \yes & \yes & \yes & \yes & \no & \no & \no & \no & \no & \no  & \no \\
    \textbf{PipeTune} & \yes & \no & \yes & \yes & \yes & \yes & \yes & \yes & \yes & \no & \no  & \yes \\
    \hline
  \end{tabular}}
\end{table*}

%% file: related.tex
%!TEX root = paper.tex

\section{Related Work}
\label{sec:related}

\begin{figure*}
    \centering
    \begin{subfigure}[b]{0.31\textwidth}
        \centering
        \includegraphics[width=\textwidth]{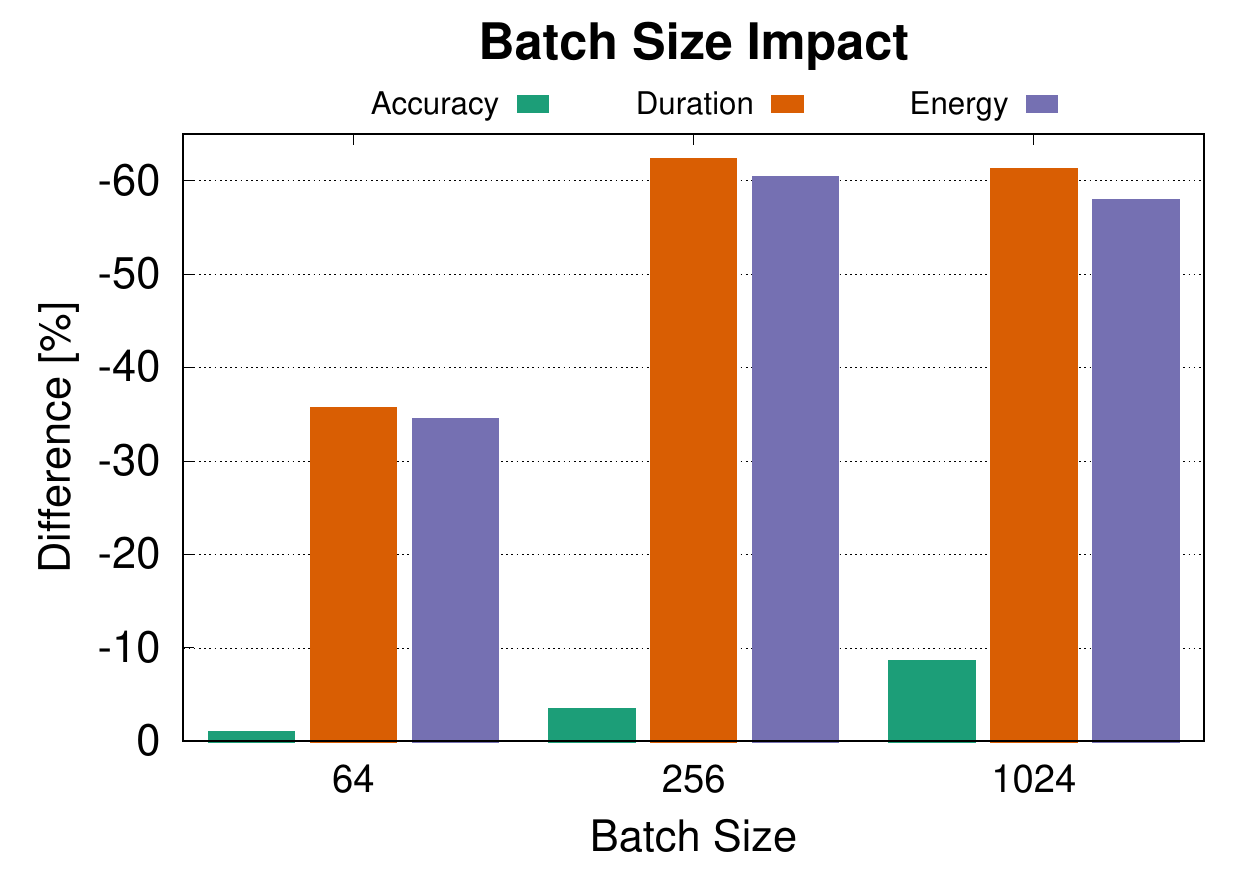}
        \caption{Batch size impact on accuracy, runtime and energy. Baseline: batch size = 32.}
        \label{subfig:batch}
    \end{subfigure}
    \hfill
    \begin{subfigure}[b]{0.31\textwidth}
        \centering
        \includegraphics[width=\textwidth]{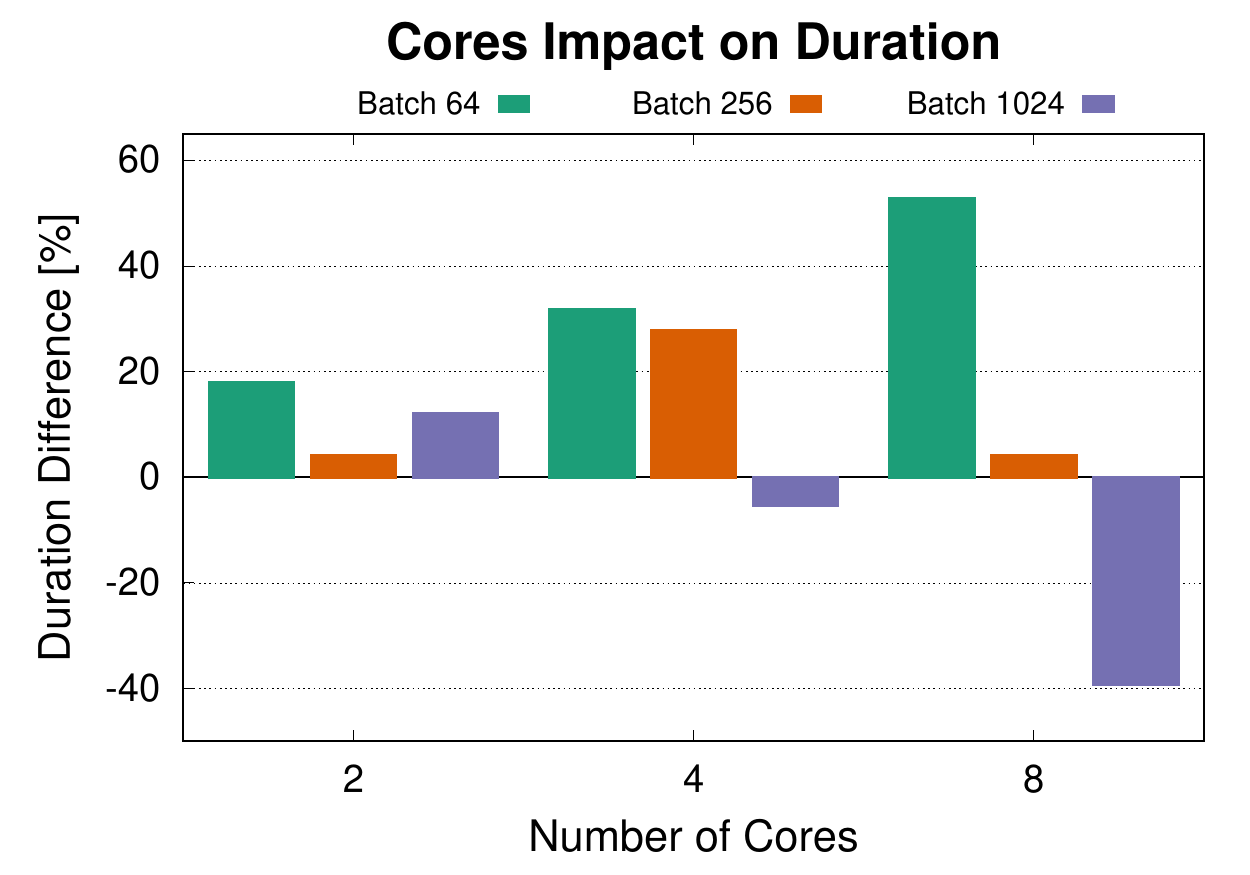}
        \caption{Cores impact on runtime per batch size. Baseline: sequential (\ie, cores = 1).}
        \label{subfig:runtime}
    \end{subfigure}
    \hfill
    \begin{subfigure}[b]{0.31\textwidth}
        \centering
        \includegraphics[width=\textwidth]{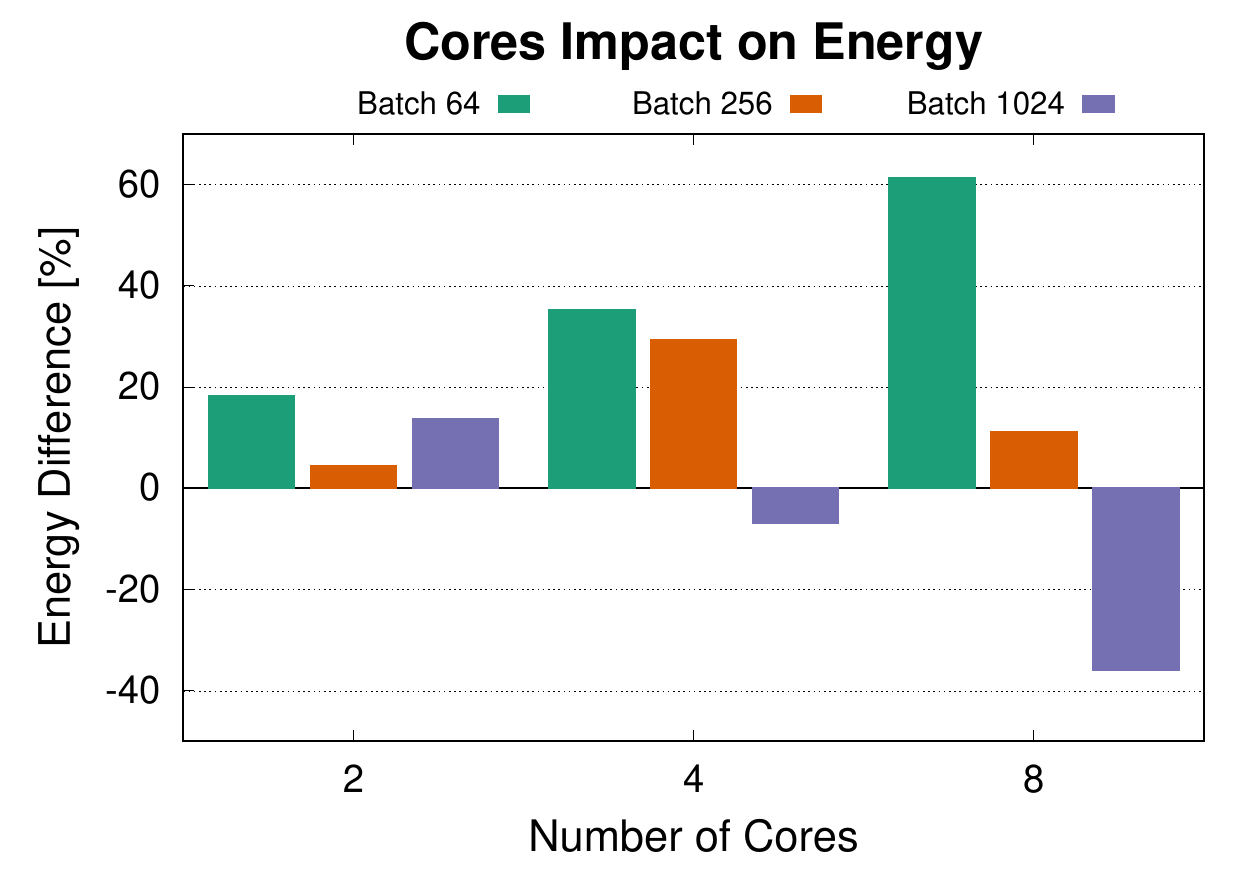}
        \caption{Cores impact on energy per batch size. Baseline: sequential (\ie, cores = 1).}
        \label{subfig:energy}
     \end{subfigure}
    \caption{Impact of hyper and system parameters on accuracy, runtime and energy training of \textsc{LeNet} and \textsc{MNIST} workload. \label{fig:parameters}}
\end{figure*}

There is a large body of work behind machine learning in general, and parameter tuning more specifically.
We survey the most prominent ones in \Cref{tab:rw}.
We distinguish between systems that support CPU or GPU processing nodes, if they can be deployed over a distributed cluster, if they support hyper or system parameters tuning.
Finally, we identify what mainstream Deep Learning frameworks are natively supported by such systems.
We distinguish between systems improving new techniques for training, others specifically optimized for hyperparameter tuning, and those focusing on system parameters tuning. 

\noindent\textbf{Improving training.} 
GRNN~\cite{DBLP:conf/eurosys/HolmesMHYW19} constructs a hybrid performance model that estimates the cost of a configuration according to the communication and computation needs. It ranks all the configurations and selects the first top-K to compile and run returning the fastest among them.

Hop~\cite{DBLP:conf/asplos/LuoLZQ19} is a heterogeneity-aware decentralized training protocol. 
It relies on a queue-based synchronization mechanism that can implement backup workers and bounded staleness in a decentralized setting.

Optimus~\cite{DBLP:conf/eurosys/PengBCWG18} uses online fitting to predict model convergence during training, and sets up performance models to estimate training speed as a function of allocated resources in each job. 
It estimates online how many more training epochs a job needs to run for convergence and how much time a job needs to complete one training epoch given a certain amount of resources. 
Speed model is computed based on a small sample set of training data, with possible combinations of parameter servers and workers. %for several steps

Orion~\cite{DBLP:conf/eurosys/WeiGGX19} performs static dependence analysis to determine when dependence-preserving parallelization is effective and map a loop computation to an optimized distributed computation schedule.
It automatically parallelizes serial imperative ML programs on distributed shared memory. 

Parallax~\cite{DBLP:conf/eurosys/KimYPCJHLJC19} combines Hyperparameter Server~\cite{li2014scaling} and AllReduce~\cite{mamidala2004efficient} architectures to optimize the amount of data transfers according to the data sparsity. 
It splits between a static phase for graphs with dense variables, and a sampling phase for fewer iterations.

PipeDream~\cite{DBLP:conf/sosp/NarayananHPSDGG19} combines inter-batch pipelining and intra-batch parallelism to improve parallel training throughput, helping to better overlap computation with communication and reduce when possible the amount of communication.

These approaches focus on optimizing the training process, and can be combined with \sys to achieve further performance gains.

\noindent\textbf{Hyperparameter tuning.} 
As the process of tuning hyperparameters is, in most cases, crucial to find the best model performance of a given application, there are many proposed approaches and tools addressing this problem. 

Astra~\cite{DBLP:conf/asplos/SivathanuCSZ19} is a framework for online fine-grained exploration of the optimization state space in a work-conserving manner while making progress on the training trials.

STRADS~\cite{DBLP:conf/eurosys/KimHLZDGX16} exposes parameter schedules and parameter updates as separate functions to be implemented by the user. A parameter schedule identifies a subset of parameters which a given worker should sequentially work on.
STRADS-AP~\cite{DBLP:conf/usenix/KimAGX19} extends STRADS to a distributed ML setting. 
These approaches leverage a runtime and API comprised of Distributed Data Structures (DDSs) and parallel loop operators.

AutoKeras~\cite{jin2019auto} enables Bayesian optimization to guide the network morphism for efficient neural network architecture search. 
The framework develops a neural network kernel and a tree-structured acquisition function optimization algorithm to efficiently explore the search space.
Similarly, Tune~\cite{liaw2018tune} provides a narrow-waist interface between training and search algorithms.

Finally, we mention two auto-tuning tools used in industry.
HyperDrive~\cite{rasley2017hyperdrive} is a package part of Azure Machine Learning which supports hyperparameter tuning.
It follows POP's scheduling algorithm which combines probabilistic model-based classification with dynamic scheduling and early stop techniques.
Amazon SageMaker~\cite{liberty2020elastic} is a fully managed machine learning service.
It supports automatic model tuning component that finds the best version of a model by running many training trials on the dataset using the algorithm and ranges of hyperparameters specified by the user. 
Google Vizier~\cite{DBLP:conf/kdd/GolovinSMKKS17} is an internal service used to optimize machine learning models and other systems. It also provides core capabilities to Google’s Cloud Machine Learning HyperTune~subsystem.

As our approach is an extension of pure hyperparameter tuning, the above mentioned systems and all others which focus on hyperparameter auto-tuning could profit from \sys.

\noindent\textbf{System parameter tuning.} 
ByteScheduler~\cite{DBLP:conf/sosp/PengZCBYLWG19} is a Bayesian optimization approach.
It specifically focuses on auto-tune tensor credit and partition size for different training models under various networking conditions. 
ByteScheduler uses auto-tune algorithms to find the optimal system related configurations.
Instead, \sys allows the user to perform hyperparameter auto-tuning and finds the best system configurations independently of this process.

AutoKeras~\cite{jin2019auto} supports a form of system parameter tuning, by means of an adaptive search strategy for different GPU memory limits.
However, instead of adapting the system parameters to the workload, as we do in \sys, AutoKeras limits the size of the neural networks according to the GPU memory.

To the best of our knowledge, \sys is the first solution that efficiently combines hyper and system parameters in a holistic manner.

%% file: background.tex
%!TEX root = paper.tex

\section{DNN Tuning: A primer}
\label{sec:background}
In this section, we discuss how hyperparameter tuning operates and explain why taking system parameters into account is beneficial.
Then, \S\ref{sec:debunk} experimentally shows the benefits of our rationale.

\subsection{Hyperparameters}

A hyperparameter is a configuration external to the model.
Its value cannot be estimated from data, it is set before the training starts, and does not change afterwards. 
Choosing the right hyperparameters during the \emph{tuning} phase is key, as the output accuracy of the trained models can vary significantly.
This phase is typically based on trial-and-error with model selection criteria.
The complexity of this phase sparked several research efforts towards its automation and autotune frameworks~\cite{jin2019auto, liaw2018tune, DBLP:conf/kdd/GolovinSMKKS17}.   
As a result, hyperparameter optimization outputs a tuple of hyperparameters that yields an optimal model which minimizes a predefined loss function on given independent data~\cite{DBLP:journals/corr/ClaesenM15}.

Typically the selection criterion considered is model accuracy. 
However, the hyperparameters values will impact model accuracy, its training time and the energy footprint. 
The former is typically related to the utility of the trained model, the latter two to its costs. 
\Cref{subfig:batch} shows this behavior by reporting the impact of varying one hyperparameter (\ie, batch size) for the training of a \textsc{LeNet} model~\cite{lecun2015lenet} on the \textsc{MNIST}~\cite{mnist} dataset. 
On the y-axis we show the measured differences for accuracy, duration and energy observed for~3 possible batch size values (\ie, 64, 256, and 1024), against a default value of 32.

We can observe how larger values of batch size achieve worse accuracy, but shorter training time and lower energy footprints. 
However, these observed trends might present considerable variations for different applications as it strongly depends on the workload and the values of the other hyperparameters. 
Therefore, these trade-offs are not trivially predicted, making it challenging to handle multidimensional selection criteria.

\begin{figure}
    \centering
    \includegraphics[scale=0.65]{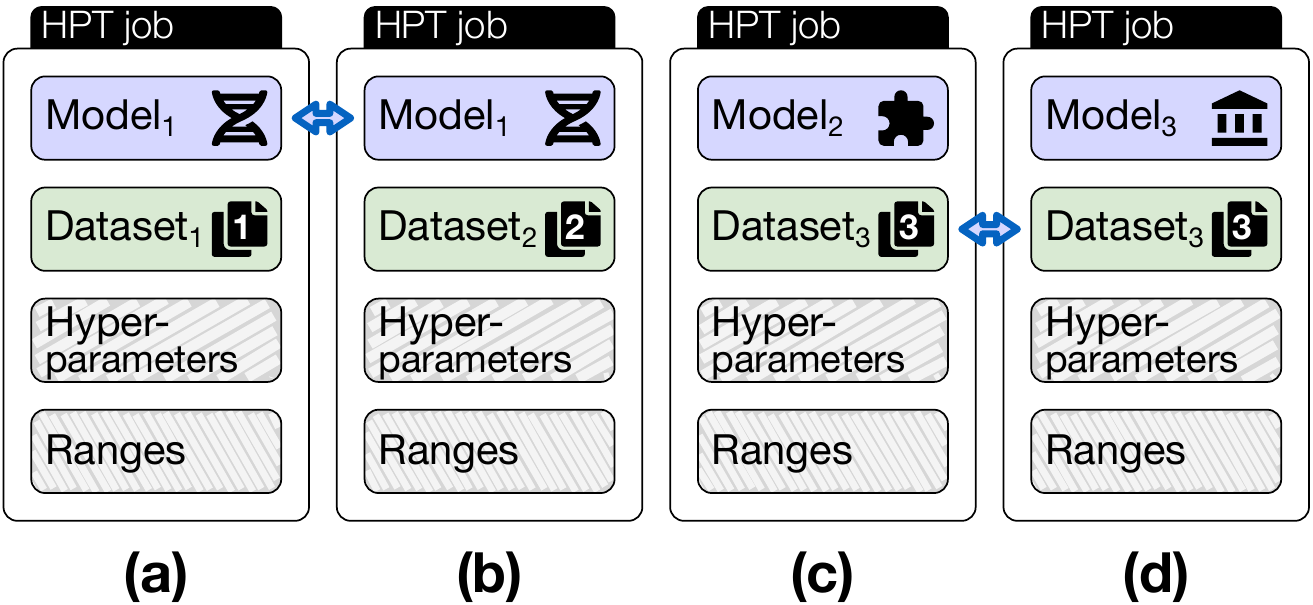}
    \caption{Workloads similarity of HPT jobs on the model and dataset level. Jobs (a) and (b) consist of the same model. Jobs (c) and (d) consist of the same dataset.}
    \label{fig:workloads}
\end{figure}

\subsection{System Parameters}

We define system parameters the configurable resources of the underlying computing infrastructure where the training will execute (\eg, memory, CPU cores, CPU frequency). 
Typically, the hyperparameter optimization fixes the same system parameters for each trial, although they might benefit from different configurations. 
To highlight this, we train again a \textsc{LeNet} model on the \textsc{MNIST} dataset.
We vary the number of CPU cores used with different batch sizes. 
\Cref{subfig:runtime} and \Cref{subfig:energy} depict our findings.
We observe how the number of cores is beneficial for larger batch size values (\eg, 1024), but not for smaller ones. 
In fact, for smaller values (\eg, 64) the runtime increases as the number of cores increases. 
This behavior is explained by the synchronous mini-batch stochastic gradient descent (SGD) algorithm used to train the neural network model.
Each $N$ iterations, SGD first computes the gradients using the current mini-batch, and then makes a single update to the weights of the neural network model. 
The \textit{batch size} hyperparameter is divided by $N$ to form these mini-batches, where $N$ is the number of cores. 
When this value is too small, the overhead of model parameters synchronization is too high and ends up slowing down the training itself. This overhead can be amortized by using techniques such as the ones implemented by Drizzle \cite{venkataraman2017drizzle} which schedules multiple iterations of computations at once, greatly reducing scheduling overheads even if there are a large number of tasks in each iteration
~\cite{SOCC2019_BIGDL}. 

Regarding the energy observations, we estimate the overall energy consumption of the cluster by calculating the trapezoidal integral of the power values collected every second during training. 
We observe a clear correlation between the energy variations (\Cref{subfig:energy}) and training runtime's gains (\Cref{subfig:runtime}). 
These observations might however vary when the tuning is applied to different set of system parameters, \eg, CPU frequency, or for different workloads.

In summary, these preliminary results show the delicate trade-offs between hyper and system parameters.
One needs to balance them all towards optimal values, such that the underlying system achieves the best training performance without compromising the model accuracy.

\begin{figure}
    \centering
        \includegraphics[width=\columnwidth]{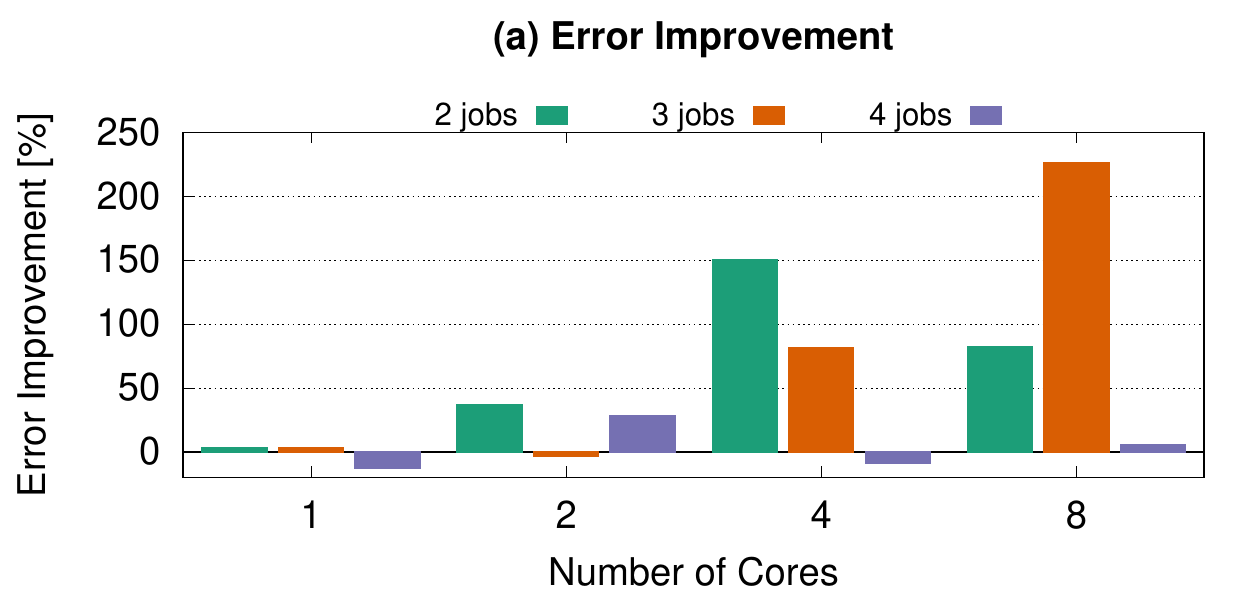}
        \includegraphics[width=\columnwidth]{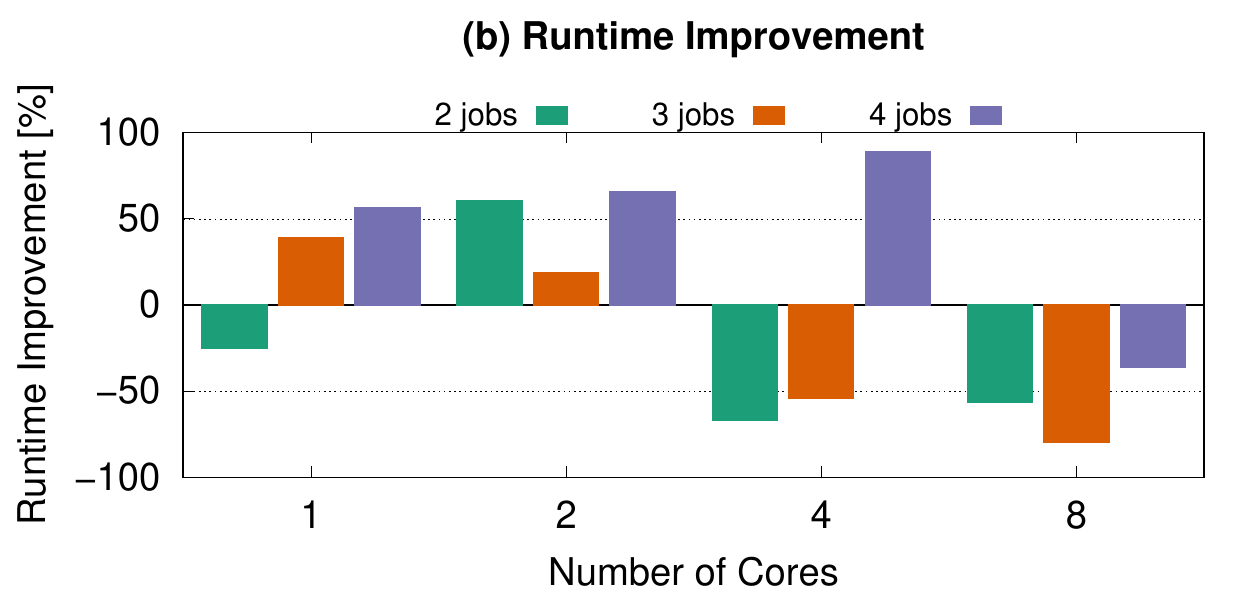}
        \caption{Characterizing \baselinesys's performance under various system conditions (\ie, system load, number of cores, and hyperparameters) during tuning.}
    \label{fig:tune_fail}
\end{figure}

\subsection{Workload}

A workload is a tuple pairing a model and dataset. 
Typically, DNN workloads are used for training (\ie, learning) or inference (\ie, prediction).
In this work, we only consider the training phase of DNN workloads.
Moreover, we assume that this training phase includes parameters tuning on top of learning the weights of the model. 
Hence, tuning a single workload consists of multiple training trials, each divided into epochs. 
Each epoch involves one forward and one backward pass of the entire input dataset. 
For ease of processing, the dataset is split into smaller batches, and each batch is propagated forward and backward once during an epoch (\ie, iteration). 
These mechanisms apply generally to all DNNs.
It is a common practice to train the same model with different datasets, as well as different models using the same dataset. 
\Cref{fig:workloads} depicts this practice.
Our approach leverages the similarity existent among such jobs to improve the tuning~performance.

%% file: motivation.tex
%!TEX root = paper.tex

\section{The "System \emph{as} Hyperparameters" Case}
\label{sec:debunk}

\begin{figure}[!t]
    \centering    
	\includegraphics[scale=0.65]{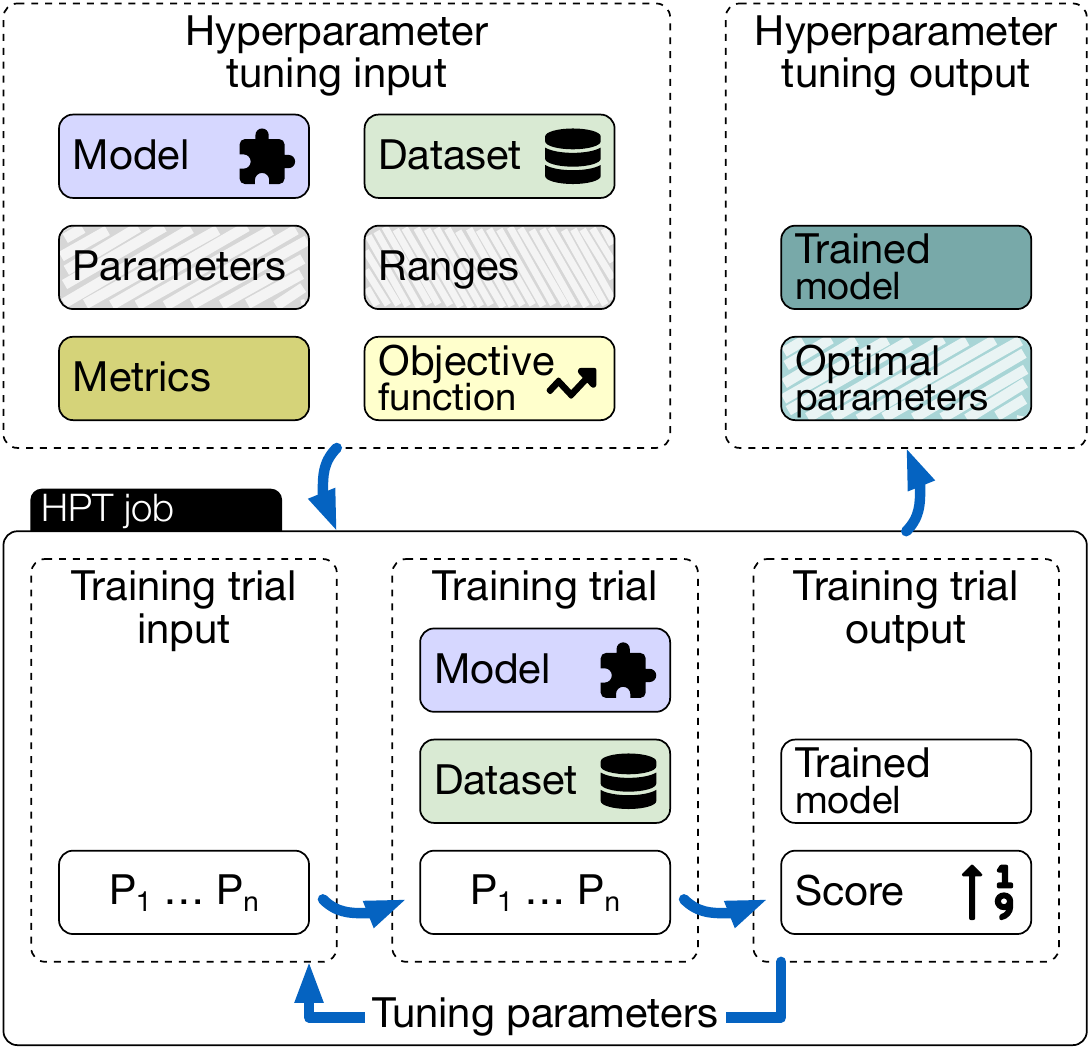}
    \caption{Hyperparameter tuning flow.\label{fig:hyperparameters_tuning}}
\end{figure}

The idea to consider system parameters as an additional set of hyperparameters is appealing.
To verify its viability, we consider a state-of-the-art hyperparameter auto-tuning system, \baselinesys~\cite{liaw2018tune}, an open-source library implemented in Python supporting an extensive list of hyperparameters optimization algorithms. 
Note that the ideas shown next are nevertheless independent of the underlying tool used for the auto-tuning process of hyperparameters.

First, we consider two versions of \baselinesys.
In V1, it is used out-of-the-box to perform hyperparameters tuning with the objective of maximizing accuracy, without taking the system parameters into account. 
In this version all trials run with the same default system parameters. 
Then, in V2, the system parameters are included in the list of parameters to be tuned. 
This second version requires the resources used by each trial to be manually controlled.
Also, the objective function must be adapted to maximize the ratio accuracy to duration, rather than restricting it to accuracy only.

\Cref{fig:tune_fail} shows the results of \baselinesys's performance characterization under various system conditions (\ie, the number of cores assigned to the tuning job and the number of jobs assigned to the same logical cores). 
We used the V2 version of \baselinesys to perform hyperparameter tuning. 
The tuning process was pinned to the same set of cores as the background jobs. 
For example, a configuration of 2 cores and 3 jobs meant a tuning job and 2 background jobs used the same 2 cores for execution. 
\Cref{fig:tune_fail} (a) illustrates the improvement in error relative to a single \baselinesys V1 job.
\Cref{fig:tune_fail} (b) is similar but shows training time improvement. 
Tuning under different system conditions significantly impacts the performance of the model being trained. 
There are only a few system configurations that yielded improvements over the baseline for error and training time. 
Some system configurations caused the tuning to trade better accuracy for faster training. 

Hyperparameter tuning without system conditions can produce less efficient models.
\Cref{tab:compare} shows the accuracy, training and tuning time achieved by different approaches for a \textsc{LeNet} model on \textsc{MNIST} dataset. 
These results show us the following. 
First, arbitrary values, if not correctly chosen, lead to both worse accuracy and training time.
Second, if the user's focus is accuracy only, then \sys's accuracy results are comparable to \baselinesys V1 however achieve in a lower tuning time.
Third, if the user's focus is both accuracy and training time, then \sys's training time results are comparable to \baselinesys V2 but with better accuracy and lower tuning time as well.

%% file: system.tex
%!TEX root = paper.tex

\section{The \sys System}
\label{sec:system}

This section presents the system design of \sys. 
We begin clarifying the problem addressed by our system~(\S~\ref{sec:problem}).
Then, we showcase its workflow (\S~\ref{sec:workflow}), the role of \sys's profiling phase (\S~\ref{subsec:profiling}), the ground-truth phase (\S~\ref{subsec:groundtruth}) and finally probing (\S~\ref{subsec:probing}).

\begin{table}[!t]
  \caption{Accuracy, training and tuning time taken by each considered approach for \textsc{LeNet} model on \textsc{MNIST} dataset.}
  \label{tab:compare}
  \rowcolors{1}{white}{gray!10}
  \resizebox{\columnwidth}{!}{
    \begin{tabular}{lccc}
    \textbf{Approach} & \textbf{Accuracy [\%]} & \textbf{Training Time [s]}  & \textbf{Tuning Time [s]} \\
    \bottomrule
    Arbitrary & 84.47 & 445 & - \\
    \baselinesys V1 & 91.54 & 272 & 4575 \\
    \baselinesys V2 & 81.76 & 187 & 4817 \\
    \sys & 92.70 & 188 & 3415
    \end{tabular}
  }
\end{table}

\subsection{Problem statement}
\label{sec:problem}

One of the first challenges of applying deep learning algorithms in practice is to find the appropriated hyperparameter values for a given workload. 
We assume that most DNN tuning jobs make use of some existing hyperparameter optimization solution. 
In the following we refer to these types of jobs as HPT~Jobs (\ie, Hyperparameters Tuning Jobs). 

A given HPT~Job takes as input a given workload, a set of parameters, its respective set of range values, an objective function and the metric of interest (\eg, accuracy, performance, energy). 
This job spawns a collection of \textit{training trials} based on the possible values of the parameters, following a given search algorithm (\eg, GridSearch, HyperBand).
Each \textit{training trial} takes as input the workload and a set of fixed values for the parameters of interest, where these values belong to their respective given ranges.
These trials can run either sequentially or in parallel depending on the setup. 
They produce a trained model and a score for the given parameters values. 
Scores correspond to the metric of interest defined by the user. The optimal set of parameters values is chosen by applying the objective function to the scores. 
\Cref{fig:hyperparameters_tuning} illustrates this process.

\begin{figure}[!t]		
     \centering		
     \includegraphics[scale=0.65]{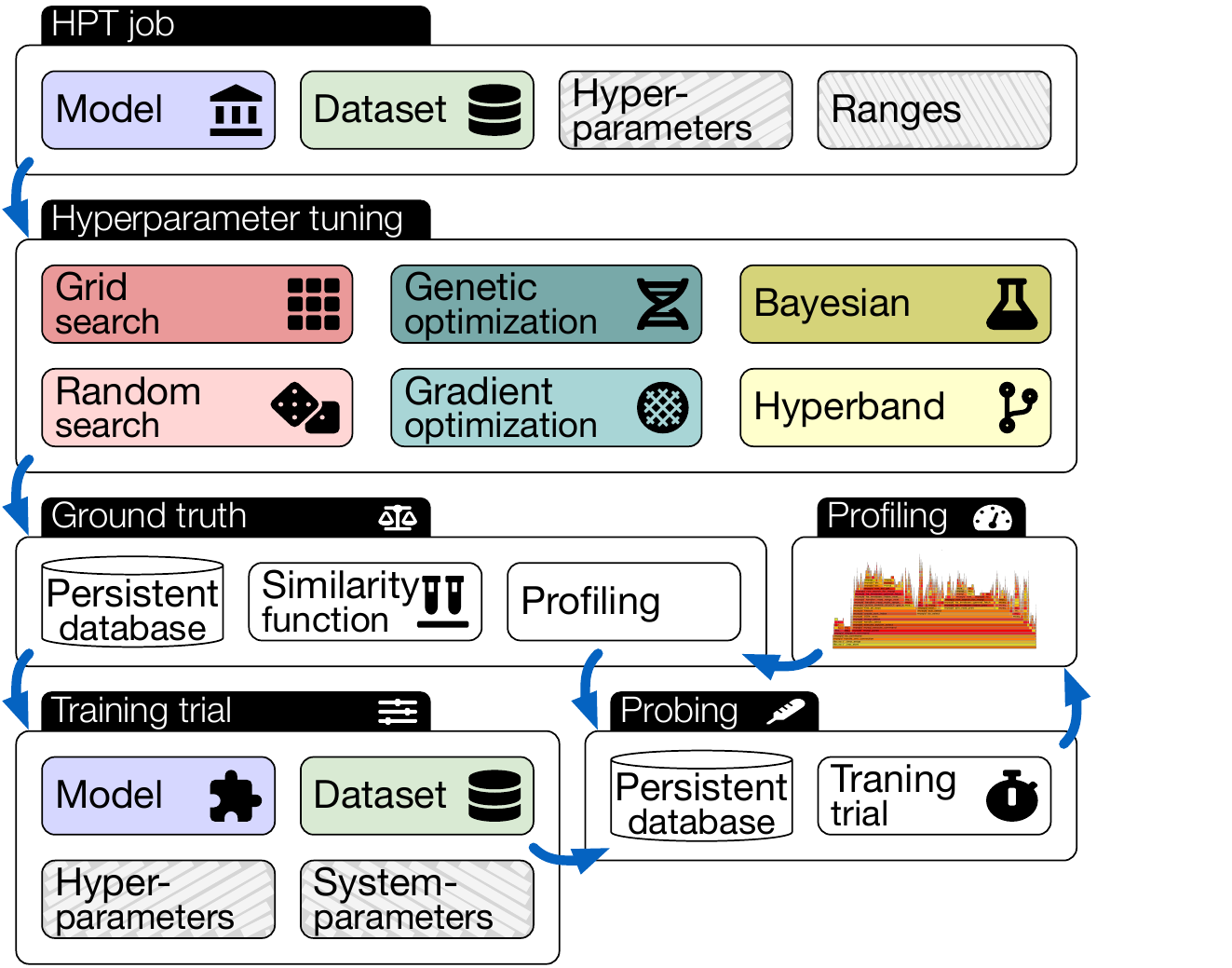}		
     \caption{\sys architecture.}		
     \label{fig:architecture}		
 \end{figure}

We consider a deep learning cluster consisting of $N$ nodes, each containing $C$ cores and $M$ GB of memory.
Note that despite a common trend to include GPUs in DNN clusters, we explicitly put aside this option.
We do this given the (rather small) nature of jobs on which we focus, for which commodity machines are sufficient for training.
HPT Jobs are scheduled in a FIFO manner.
We categorize these jobs in the following two main types: Type-I: tuning the same model for different datasets (\eg, recommendation engines), and Type-II: tuning different models for the same dataset (\eg, computer vision).

Both types of tuning jobs can still be divided into two sub-types: \emph{(a)}~same set of hyperparameters and ranges, and \emph{(b)}~same set of hyperparameters but different ranges. 
Each job, independent of its category, performs the earlier described tuning process from scratch. 
\textbf{A key observation is that these jobs could benefit from previously computed results for other jobs in the same category to converge faster}. 
Moreover, training trials spawned by the same HPT Job run all with the same system parameters even though they might require different resources configuration.

Another major limitation of the currently available approaches to hyperparameter auto-tuning is that only a single objective metric can be specified.
This means that for a given HPT~Job, one could choose to optimize either accuracy or performance, but not both simultaneously.

In summary, our problem's input consists of an HPT Job with the objective of achieving either maximum accuracy, or maximum accuracy with minimum training time. 
The former must output the best possible hyperparameters leading to the highest accuracy, independent of training time. 
For the latter, a combination of optimal hyper and system parameters is expected which leads to the highest accuracy and lowest training time. 
Note that for both scenarios, a shorter tuning times is beneficial, as allowed by our approach.

\begin{algorithm}[!t]
\caption{\sys algorithm.}
\label{alg:pipetune}
\SetKwProg{Train}{Function}{:}{end}
\Train{train(model, data, hyperparameters)}{
    job = \textit{\textbf{async} model.train(data, hyperparameters)}\;
    \textit{\textbf{async} tuneSystem(job)}\;
    job.wait()\;
    \textbf{return} model\;
}

\SetKwProg{SysTune}{Function}{:}{end}
\SysTune{tuneSystem(model, data)}{
    profile = \textit{getProfile(job)}\;
    (score, config) = \textit{getSimilarity(profile)}\;
    \If{score > threshold}{
        \textit{setSystemParameters(config)}\;
    }
    \Else{
        \ForEach{$sp_v \in \textit{systemParameters}$}{
        \textit{setSystemParameters($sp_v$)}\;
        wait until epoch finishes\;
        add collected metrics to $m$\;
        }
        bestConfig = find best config in $m$\;
        \textit{setSystemParameters(bestConfig)}\;
  }
}
\end{algorithm}

\subsection{\sys Workflow}
\label{sec:workflow}

\Cref{fig:architecture} depicts the architecture components of \sys design and the main workflow.
While training hyperparameters, a \textit{trial} is a single training run with a fixed initial hyperparameter configuration. 
In order to find the best values for a given set of hyperparameters, the system executes a collection of trials, supervised by a given tuning library (\eg, Vizier, Tune) and using one of the supported trial scheduling algorithms (\eg, GridSearch, HyperBand).

\sys enhances the tuning of system parameters following a pipelined parallelism approach. 
That is, within each trial, a collection of sub-trials is executed, with the goal of  defining the best system configurations for a given optimization function and metric of interest.
This sub-trial consists of varying the system configuration on the epoch level and monitoring the system itself as well as the metrics of interest.
The execution of sub-trials is controlled by \sys, which may also rely on different underlying scheduling algorithms.

\Cref{alg:pipetune} details the pipelined approach.
Function \textit{train} (lines~1-5) is executed during a trial for a given workload (\ie, model and dataset). 
After initiating the model training using the hyperparameter configuration given for that trial, \textit{tuneSystem} (line~3) is invoked asynchronously. 

The \textit{profiling} phase (lines 7) is initiated for this given trial with the objective of characterizing the workload properties and its systems requirements. 
This process is done at the granularity of epochs for the currently running trial.
We rely on kernel performance counters (\eg, cpu cycles memory stores, instructions) to gather hardware events corresponding to low-level metrics of the underlying system.

Once this \textit{profiling} phase is over, its outcome is used as input to a \textit{ground truth} phase. 
This process consists of applying a similarity function (line~8) on the job's profile.
This is done to reuse optimal configurations known by the system for other jobs with similar characteristics. 
If the score of this similarity function is within a specific confidence level (line~9), then the optimal known configurations are applied (line~10) and no further system metric trials are required. 
However, if the score does not cross the threshold, a new \textit{probing} phase starts, searching the optimal system configurations for that~trial. 

The probing requires each system configuration to be applied for a different epoch, following a given scheduling algorithm. 
We collect several meaningful metrics (\eg, runtime, energy) plus low-level metrics (\eg, hardware events). 
Then the optimization function is applied over these metrics (line 16) to identify the overall best system configuration. 
This process consists of iterating over the collected values for each tuple of system parameters, looking for the one which best fits the optimization function (\eg, shortest runtime, lowest energy consumption). 
The complexity of this search is $O(n)$, where $n$ is the number of distinct system parameters considered.
Finally, the configuration identified as optimal is applied for the remaining iterations (line~17) and saved for further improving of the \textit{ground truth} phase. 

\subsection{Profiling}
\label{subsec:profiling}

The profiling component leverages hardware performance counters to collect low-level events of the system during the applications execution time. 
After an initial experiment campaign, we gathered a comprehensive list of such events.
As the number of events collected per time unit is limited by the number of actual hardware counters of the CPU, we filter out highly correlated as well as unsupported events. 
As result, our prototype deployed on x86 architectures current considers 58 measurable events, most of them being Performance Monitoring Unit (PMU) hardware events (\eg, branch-instructions, cache-misses, cpu-cycles, mem-loads), reported by Linux's \textit{perf} (v4.15.18).
Although we have filtered the list of possible events to be collected, common Intel processors have only 2 generic and 3 fixed counters. Generic counters can measure any events while fixed counters can only measure one event. 

When there are more events than counters (as it is in our case), then the kernel uses time multiplexing to give each event a chance to access the monitoring hardware. 
When this happens, an event might miss a measurement.
If this happens, its occurrences are recomputed once the run ends, based on total time enabled \emph{vs} time running~\cite{perf}, with:
\[final\_count = raw\_count * time\_enabled/time\_running.\]
This provides an estimate of what the count would have been, had the event been measured during the entire run. 

Considering that the output value is not an actual count, depending on the workload, there might be blind spots which can introduce errors during scaling.
Although we profile workloads at the epochs granularity, each epoch runs for at least a few minutes and we measure the events of interest every second. To mitigate the potential profiling errors, we store the average of results during each epoch's time window.

\subsection{Ground Truth}
\label{subsec:groundtruth}

During this phase, new incoming HPT Jobs exploit the ground truth results from historical data collected during the previously completed jobs with similar system characteristics, to accelerate their system-parameter tuning phases.
Our design allows the similarity function to be pluggable, and while we do settle on k-means~\cite{wagstaff2001constrained} in the current implementation, \sys allows to easily switch to alternative techniques.

The implementation of \textit{ground truth} is done as a separate module which is used by \sys. 
In this module, the user can point to a pre-trained similarity function for a warm start or let the system build a new one from scratch. 
For this, our currently implementation relies on the \textit{scikit-learn} machine learning library for Python \cite{pedregosa2011scikit} which already supports several clustering algorithms (\eg, affinity propagation, mean-shift, DBSCAN, OPTICS, Birch).
The exhaustive list of supported models are then inherited by \sys and could be easily used as alternative similarity functions.

Regarding the currently used model (\ie, k-means), it is trained over the low-level system metrics collected during the profiling phase.
The datasets are then partitioned into $k=2$ groups (\ie, model and dataset).
Extensions to other values of $k$, as well as to other similarities dimensions (\eg, hyperparameters, ranges) are left for future work.

\Cref{fig:clsuters} shows clustering results using k-means grouped by model and dataset labeled with their respective cluster's labels. 
We can observe that the majority of data fits into Type-I and Type-II are labeled as \textit{cluster1} and \textit{cluster2}, respectively. 
This result supports our assumptions regarding workloads similarities and shows that the chosen profiling technique can also capture the implicit characteristics of each workloads. 
Finally, it shows that the clustering algorithm utilized can identify the similarities present in those characteristics and efficiently cluster them.

\subsection{Privacy concerns} 

Although the \textit{ground truth} component of \sys makes use of historical data, it does not require any information regarding the users' workloads (\ie, model or dataset). 
Instead, this process relies entirely on system events collected using the hardware performance counters. 
This profiling based on low level metrics allows \sys to characterize the applications while preserving user data privacy (\eg, user parameters like model and dataset are not revealed).
We assume that potential data, model and parameters similarities between workloads will affect the collected metrics in the same ways and therefore also be reflected in the similarity function. 
The results observed in \Cref{fig:clsuters} supports this~assumption.

\subsection{Probing}
\label{subsec:probing}

The probing phase profiles a given set of workloads in different system conditions, in order to collect sufficient data for a warm start of the \textit{ground truth} component. 
In practice, the \textit{ground truth} model is refined as the similarity of the incoming jobs with the historical data of the system starts to decrease. 
When this happens, we launch a grid search on the system-parameters at the epoch granularity, yet other search strategies are possible. In this case, the tuning of system parameters for the current job is performed directly on the analytical data collected. Moreover, this collected data is saved to be taken into account once re-clustering is applied.

We decide upon the necessity to launch a new probing or not for a given workload based on the similarity score outputted from the ground truth phase.
When using k-means, the threshold matches the distance from the new set of data points to their current cluster's centroid. 
The distance is compared against the models' inertia, to measure the reliability of the prediction, or else if a re-clustering is needed.

%% file: implementation.tex
%!TEX root = paper.tex

\section{Implementation}
\label{sec:implementation}

\sys is implemented in Python (v3.5.2) and it consists of 947 LOC. 
We leverage two open-source projects, namely Tune and BigDL.
Tune~\cite{liaw2018tune} is a Python library for hyperparameter search, optimized for deep learning and deep reinforcement learning~\cite{li2018deep}. 
Tune provides several trial schedulers based on different optimization algorithms. 
While we select HyperBand for the reminder of this work, Tune allows to switch among the available ones, as well as to implement new ones.
As a consequence, \sys indirectly supports all its hyperparameter optimization algorithms.

The training applications are executed by BigDL~\cite{SOCC2019_BIGDL}, a distributed deep learning framework on top of Apache Spark. 
BigDL supports TensorFlow and Keras, hence \sys supports models defined using such frameworks. The Ground Truth module is based on a battle-tested k-means implementation openly available in the \textit{scikit-learn} machine learning library for Python \cite{pedregosa2011scikit}.

Finally, as storage backend, we leverage  \texttt{InfluxDB} (v1.7.4), an open-source time series database. 
It offers a convenient \sloppy{\texttt{InfluxDB-Python}} client for interacting with InfluxDB which we use to query information regarding the collected system metrics. \sys is released as open-source~\footnote{https://github.com/isabellyrocha/pipetune}.

%% file: evaluation.tex
%!TEX root = paper.tex

\begin{table}[!t]
  \caption{Workloads used for experiments.}
  \label{tab:workloads}
  \rowcolors{1}{white}{gray!10}
  \resizebox{\columnwidth}{!}{
  \begin{tabular}{rllccc}
    & \multicolumn{1}{l}{\textbf{Model}} & \textbf{Dataset}  & \textbf{Datasize} & \textbf{Train Files} & \textbf{Test Files}\\
    \bottomrule
    \cellcolor{white} & \textsc{LeNet5} & \textsc{MNIST} & 12 MB & \num{60000} & \num{10000} \\
    \multirow{-2}{*}{\textbf{Type-I}} & \textsc{LeNet5} & \textsc{Fashion-MNIST} & 31 MB & \num{60000} & \num{10000}\\
    \bottomrule
    \cellcolor{white} & \textsc{CNN} & \textsc{News20} & 15 MB & \num{11307} & \num{7538} \\
    \multirow{-2}{*}{\textbf{Type-II}} & \textsc{LSTN} & \textsc{News20} & 15 MB & \num{11307} & 7538 \\
    \bottomrule
    \cellcolor{white} & \textsc{Jacobi} & \textsc{Rodinia} & 26 MB& 1650 & 7538 \\
    \textbf{Type-III} & \textsc{SPK-means} & \textsc{Rodinia} & 26 MB & 1650 & 7538 \\
     \cellcolor{white} & \textsc{BFS} & \textsc{Rodinia} & 26 MB& 1650 & 7538
  \end{tabular}}
\end{table}

\section{Evaluation}
\label{sec:evaluation}

This section presents our in-depth evaluation of \sys using real-world datasets. 
Our main findings are:

\begin{enumerate}
    \item \sys achieves significant tuning speedups without affecting model performance (\ie, accuracy);
    \item By speeding up the tuning process, we also have a more energy efficient approach, not only due to the runtime reduction but also because of the more efficient utilization of system resources;
    \item The proposed approach is sensitive to varying system loads as this is also reflected on the events used to profile and our system adapts on a fine granularity (\ie, epochs level).
\end{enumerate}

\subsection{Experimental Setup}
\label{subsec:setup}

\begin{figure}[t]
    \centering
    \includegraphics[scale=0.65]{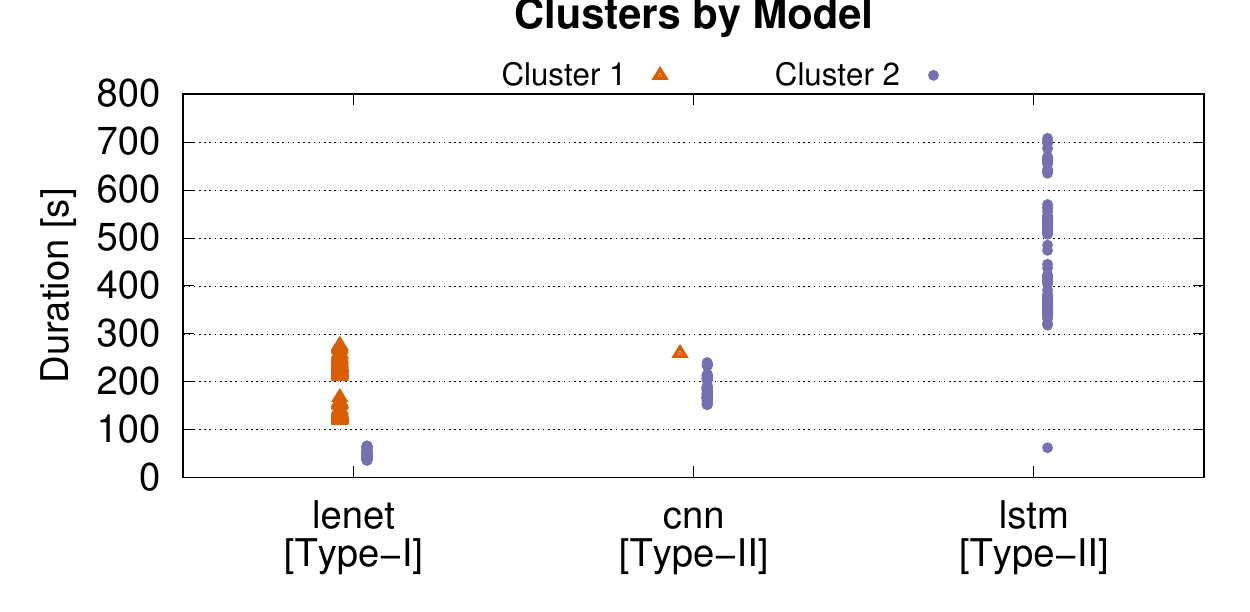}
    \includegraphics[scale=0.65]{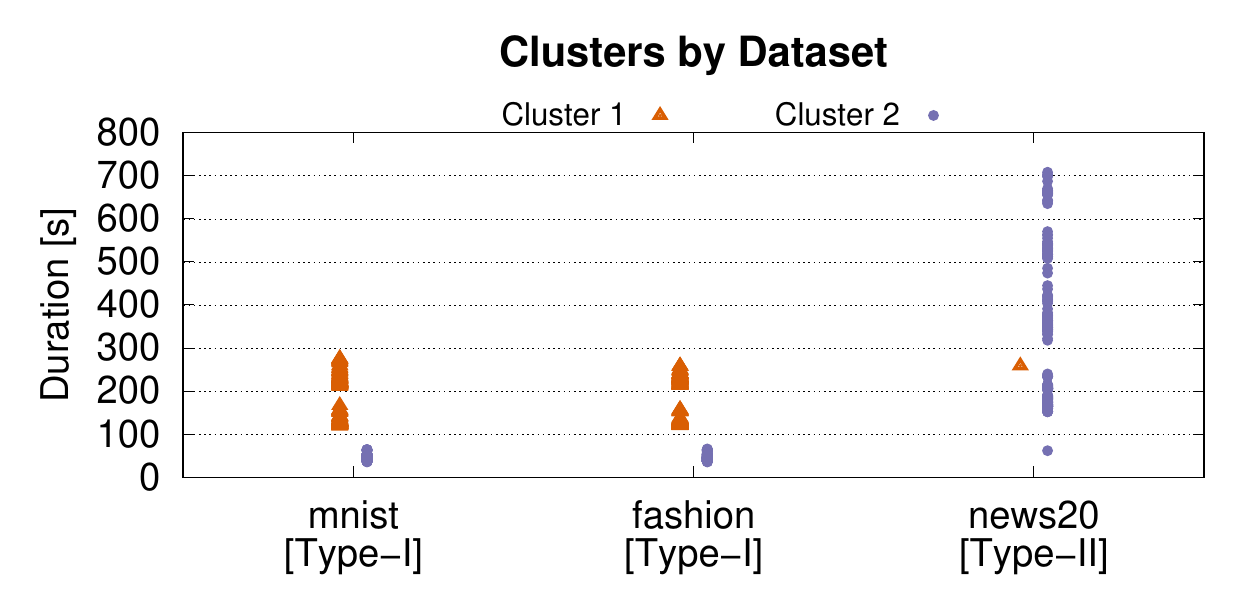}
    \caption{Clustering results grouped by workload type.}
    \label{fig:clsuters}
\end{figure}

\subsubsection{Testbed} We deploy our experiments using Type-I and Type-II workloads on a cluster of 4 quad-socket Intel E3-1275 CPU processors with 8 cores per CPU, 64 GiB of RAM and 480 GB SSD drives. Experiments involving Type-III workloads are deploy on a single node containing an Intel E5-2620 with 8 cores, 24 GB of RAM and a 1 TB HDD.
All machines run Ubuntu Linux 16.04.1 LTS on a switched 1 Gbps network. 
Power consumptions are reported by a network connected LINDY iPower Control 2x6M Power Distribution Unit~(PDU), which we query up to every second over an HTTP interface to fetch up-to-date measurements for the active power at a resolution of 1W and 1.5\% precision.

\subsubsection{Workloads} We consider 7 state-of-the-art deep learning workloads for image classification, LLC-Cache computational sprinting and natural language processing. 
Table~\ref{tab:workloads} summarizes their~details.

\textsc{LeNet5}~\cite{lecun2015lenet} is a convolutional network for handwritten and machine-printed character recognition. 
Convolutional Neural Networks (CNNs)~\cite{mikolov2010recurrent} are a special kind of multi-layer neural networks, trained via back-propagation. 
CNNs can recognize visual patterns directly from pixel images with minimal preprocessing. 
Long Short-Term Memory (LSTMs)~\cite{gers1999learning} are artificial Recurrent Neural Networks (RNNs) architectures used to process single data points (such as images, connected handwriting recognition and speech recognition), as well as sequences of data (\ie, speech,  videos).
Finally, Jacobi is a differential numerical solver, BFS is breath-first-search and spk-means is k-means implemented on top of Spark framework.

The \textsc{MNIST} dataset~\cite{mnist} of handwritten digits has a training set of \num{60000} examples, and a test set of \num{10 000} examples. 
The digits have been size-normalized and centered in a fixed-size image.
\textsc{Fashion-MNIST} dataset~\cite{xiao2017/online} is a dataset of article images consisting of a training set of \num{60 000} examples and a test set of \num{10000} examples. 
Each example is a 28x28 grayscale image, associated with a label from 10 classes. 
\textsc{Fashion-MNIST} shares the same image size and structure of training and testing splits as the original MNIST dataset. 
The \textsc{News20} dataset~\cite{news20} is a collection of \num{20 000} messages collected from 20 different netnews newsgroups.
We sample uniformly at random \num{1000} messages from each newsgroup, and we partition them by name.
The \textsc{Rodinia} Benchmark Suite ~\cite{rodinia} is a collection of profiling short-term resource allocation (\ie, computational sprinting) policies which targets heterogeneous computing platforms with both multicore CPUs and GPUs.
These workloads have the objective to classify or predict the original data reserved for testing purposes.

\begin{figure}[t]
    \centering
    \includegraphics[scale=0.65]{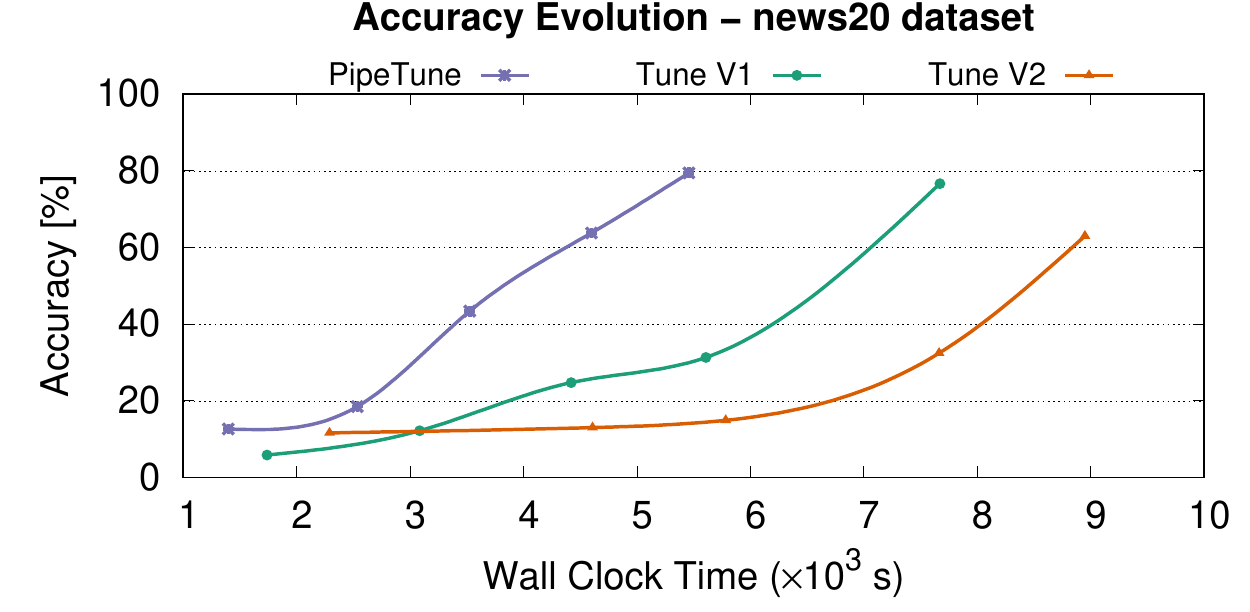}
    \caption{Accuracy convergence.}
    \label{fig:syscompareacc}
\end{figure}

\begin{figure}[t]
    \centering    \includegraphics[scale=0.65]{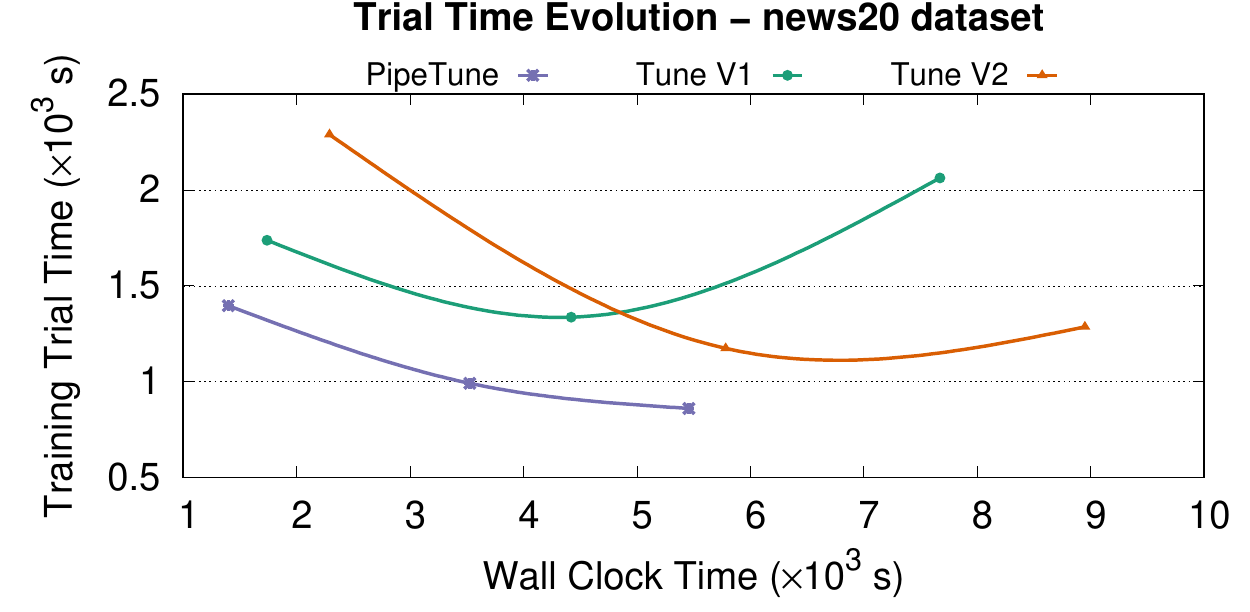}
    \caption{Training trial time convergence.}
    \label{fig:syscomparetrialtime}
\end{figure}

\begin{figure*}[!t]
    \centering
    \resizebox{\textwidth}{!}{\includegraphics[scale=0.7]{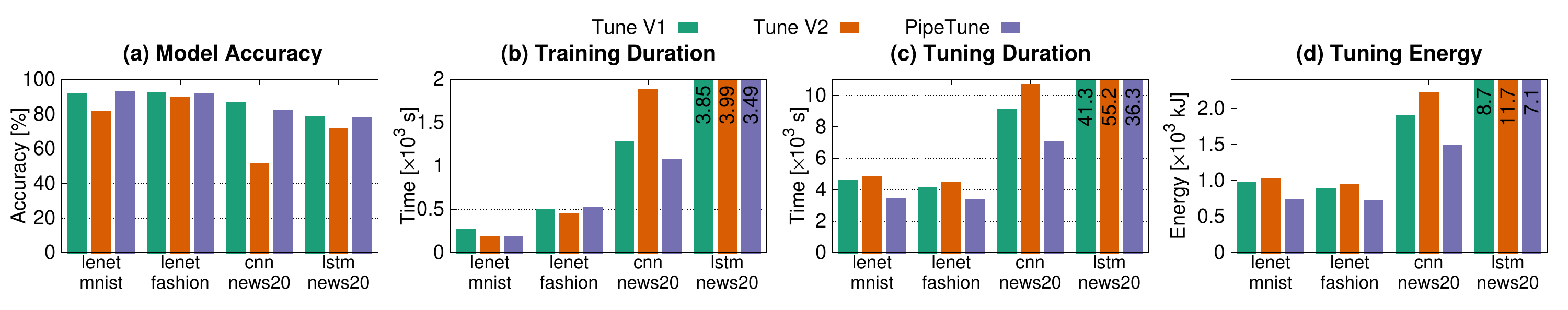}}
    \caption{Evaluation of \sys's accuracy, performance and energy consumption for Type-I and Type-II Jobs.}
    \label{fig:offline}
\end{figure*}

\begin{figure*}[!t]
    \centering
    \resizebox{\textwidth}{!}{\includegraphics[scale=0.7]{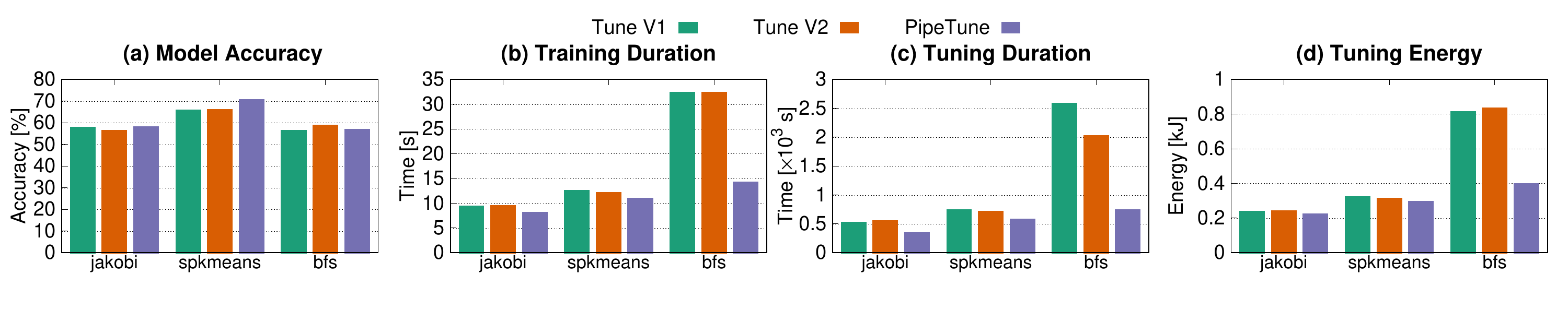}}
    \caption{Evaluation of \sys's accuracy, performance and energy consumption for Type-III Jobs.}
    \label{fig:offline_tman}
\end{figure*}

\subsubsection{Hyperparameters} 
There are several potential hyperparameters to tune.
For practical reasons, in our evaluation we select the 5 described below. 
Note that their recommended range is typically application-driven, and we settle on specific values without however generalizing for any workload.
 
\begin{enumerate}[leftmargin=10pt]
    \item \textbf{Batch size.} Number of samples to work through before updating the internal model parameters. Large values for batch size have a negative effect on the accuracy of network during training, since it reduces the stochasticity of the gradient descent. Range: [32 - 1024].
    \item \textbf{Dropout rate.} Dropout randomly selects neurons to be ignored during training. Dropout layers are used in the model for regularization (\ie, modifications intended to reduce the model's generalization error without affecting the training error). The dropout rate value defines the fraction of input to drop to prevent overfitting \cite{molchanov2017variational}. Range: [0.0 – 0.5].
    \item \textbf{Embedding dimensions.} Word embeddings provide a mean of transfer learning. This mechanism can be controlled by having word vectors fine-tuned throughout the training process. Depending on the dataset size on which word embeddings are being refined, updating them might improve accuracy~\cite{DBLP:conf/swisstext/AghaebrahimianC19}. Range: [50 – 300].
    \item \textbf{Learning rate.} Rate at which the neural network weights change between iterations. A large learning rate may cause large swings in the weights, making impossible to find their optimal values. 
	Low learning rates requires more iterations to converge. Range: [0.001 - 0.1].
    \item \textbf{Number of epochs} Number times that the learning algorithm will work through the entire training dataset. Typically, larger number of epochs yields in longer runtimes but also higher training accuracy. However, the number of epochs required to achieve a given minimum desired accuracy depends on the workload. Range: [10 - 100].
\end{enumerate}

\subsubsection{System Parameters} For the purpose of this evaluation, we restrict the list of parameters to number of cores and memory. However, the same mechanisms can be applied to any other parameter of interest (\eg, CPU~frequency, CPU voltage).
In our cluster, the ranges of valid values for system parameter tuning are [4 - 16] and [4 - 32] (GB) for for number of cores and memory, respectively. 

\subsubsection{Baselines}

\noindent\textbf{Baseline I: hyperparameters tuning.} Our first baseline system (\ie \baselinesys V1) uses the tuning of hyperparameters ignoring any system parameter. We rely on HyperBand for the parameter optimization with the objective function set to maximize accuracy.

\noindent\textbf{Baseline II: system and hyper parameters tuning.} We further compare against \baselinesys V2, where we include the list of system parameters to be considered in the list of parameters to be tuned by the HyperBand algorithm. We also include the training duration as part of the optimization function which in this baseline is set to maximize the ratio accuracy to duration (details in \S~\ref{sec:debunk}).

\subsection{Convergence Evolution}

In order to build our initial similarity model we rely on profiling data of the workloads described in Table~\ref{tab:workloads}. For each workload, we vary the system configurations as follows. 
Memory allocation can be 4GB, 8GB, 16GB, and 32GB. 
The total number of cores that could be allocated were 4, 8, or 16. Finally, batch size could take the values 32, 64, 512, or 1024. 
In total, this sums up to 48 different configurations for each workload. 
There is no reason to expect variations in the data collected from different training instances using the exact same parameters. 
However, we repeat this process twice for each configuration to make sure that the achieved model is not affected by potential unseen variations.

We begin our evaluation by analyzing the convergence trajectory of \sys compared to \baselinesys V1 and \baselinesys V2. 
\Cref{fig:syscompareacc} illustrates the accuracy evolution of the training trials over the tuning time of a \textsc{CNN} model on the \textsc{News20} dataset. 
We observe that \sys converges to an accuracy value comparable to \baselinesys V1 but at a much faster rate. 
For instance, \sys reaches a 60\% accuracy after approximately \num{4500} seconds. On average our approach is $1.5\times$ and $2\times$ faster than \baselinesys V1 and \baselinesys V2, respectively.

The training time achieved shows similar behavior (see \Cref{fig:syscomparetrialtime}).
Interestingly, \baselinesys~V1 performs worse than \baselinesys~V2. 
Since \baselinesys~V1 optimizes only for accuracy, the most accurate model not necessarily achieves the shortest training time. 
On the other hand, as \baselinesys V2 optimizes for the ratio accuracy to performance, the accuracy achieved might not be the highest possible.
However, the training time in the given configurations might be lower (which is exactly what happens in this instance of the problem). 
Finally, we observe that \sys consistently presents shorter trial times than the other two approaches during the entire tuning~process.

\subsection{Single-Tenancy}
\label{sec:performance}
We now consider a single-tenancy scenario, and assume each HPT Job runs in a dedicated cluster, where the required resources demanded by the system parameters are available and exclusive for a given tenant.
This prevents interference caused by other jobs co-located on the same cluster.
However, as a given HPT~Job spawns several \textit{training trials} asynchronously, the cluster still remains shared among these sub jobs. 
We evaluate how \sys performs in such stable setting, comparing it against \baselinesys V1 and \baselinesys V2, for all the~workloads.

\noindent\textbf{Comparison with baseline.} \Cref{fig:offline} presents the results of model accuracy, training and tuning runtime, and overall cluster energy consumption of offline HPT~Jobs for the different workloads described in Table~\ref{tab:workloads}.

\Cref{fig:offline}~(a) presents the accuracy results. 
We can observe that the accuracy of \sys is not affected by the performance optimization.
In fact, results are on par with  \baselinesys~V1, where hyperparameters tuning is done with the only objective of maximizing accuracy. 
As expected, \baselinesys~V2 decreases accuracy up to~43\%, since the objective function no longer tries to optimize accuracy but also takes the runtime into~account. 

\begin{figure}[!t]
    \centering
    \includegraphics[scale=0.68]{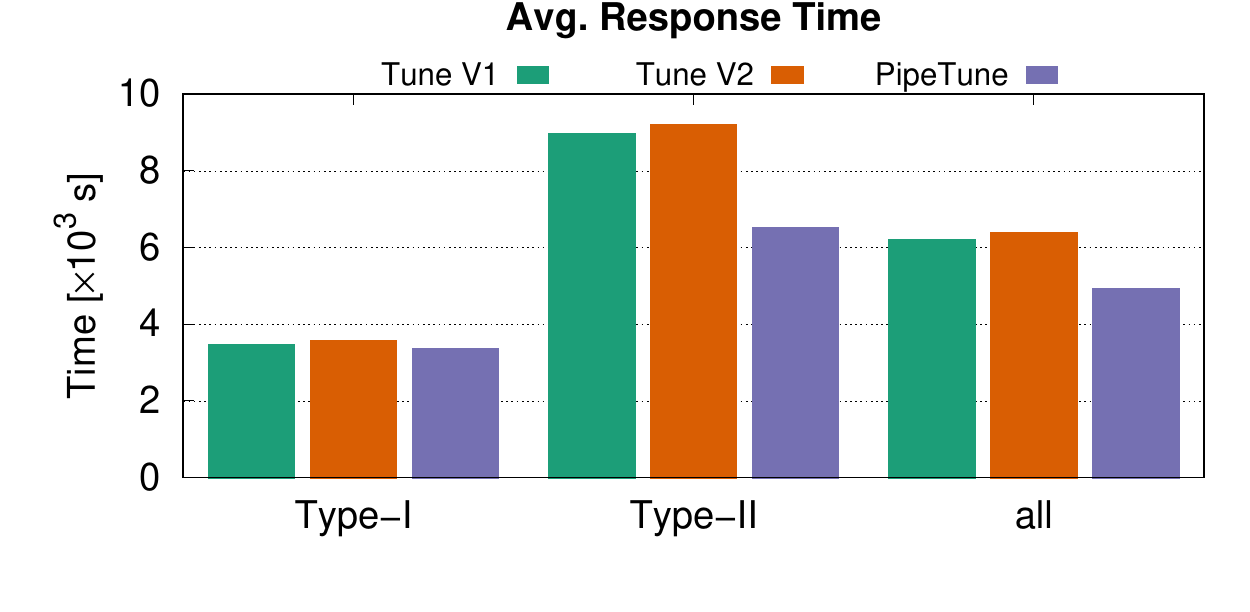}
    \caption{Average response time for Type-I and Type-II Jobs considered independently and all together.}
    \label{fig:responsetime}
\end{figure}

\Cref{fig:offline}~(b) shows the training time of the achieved model. 
In this case, \sys presents comparable results to the baseline. 
In fact, we observe up to $1.7\times$ speed-up in comparison with \baselinesys~V2 which focuses exactly in reducing training runtime. We observe that \baselinesys~V2 increases tuning duration by up to 18\% when compared to \baselinesys V1. 
This happens for the following two reasons. 
First, the search space of \baselinesys~V2 is larger than of \baselinesys~V1, as it includes the system-parameters.
Second, the optimization function consists of accuracy and runtime together. 
These two reasons make it harder for the search algorithm to find the optimal set of configurations, hence longer tuning times are observed. 

On the other hand, \sys reduces tuning runtime by at least 18\% when compared against \baselinesys~V1, as shown in \Cref{fig:offline}~(c). 
This performance gain is obtained because the search space and optimization function remains the same, and at the same time \sys finds and applies during runtime the optimal system configurations for each trial.
Moreover, all the additional steps introduced by \sys are done in parallel, without impacting the hyperparameters tuning process.

\Cref{fig:offline}~(d) reports the energy results.
The overall energy consumption of the cluster is directly affected both by the performance decays and gains. 
Compared against \baselinesys V1, we observe up to 22\% energy increase for \baselinesys V2 and up to 29\% energy decrease for \sys.

\begin{figure}[!t]
    \centering
    \includegraphics[scale=0.68]{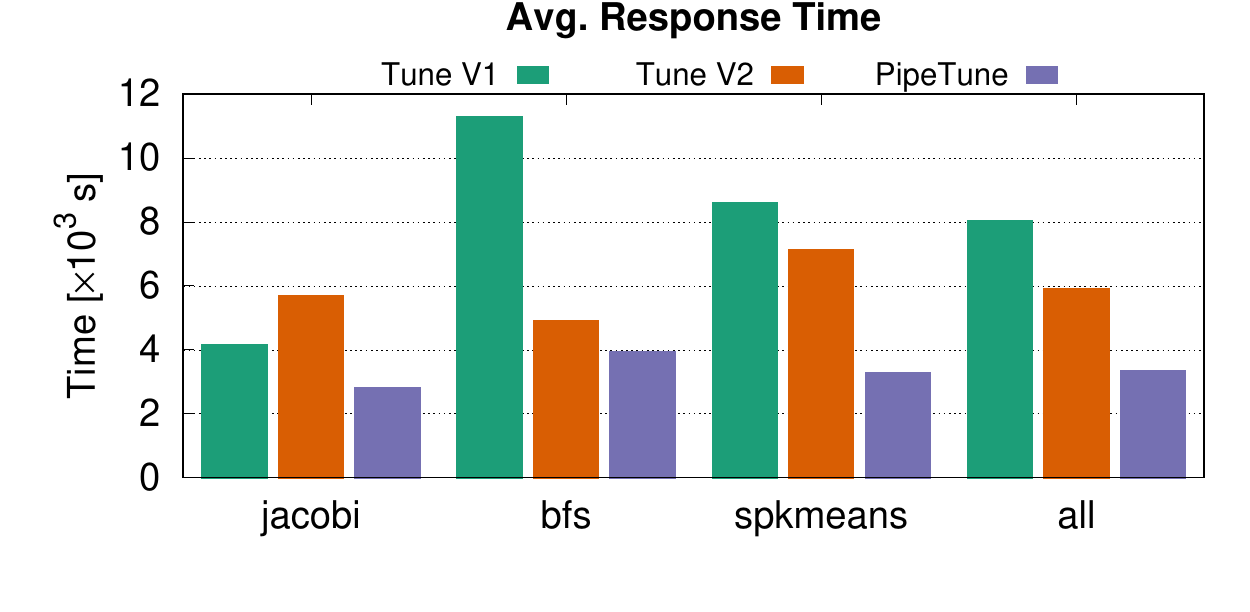}
    \caption{Average response time for Type-III Jobs.}
    \label{fig:online_type3}
\end{figure}

\Cref{fig:offline_tman} compares \baselinesys V1, \baselinesys V2 and \sys on a single node. The Type-III workloads used in these experiments have shorter epochs and each a different CNN model. Previous experiments deploy \sys on workloads with epochs lasting minutes. Long epochs work in favor of \sys since low-overhead profiling is performed across the first couple of epochs to classify new workloads. Therefore, next we perform an extra analysis on Type-III Jobs which present this more challenging setup for \sys to observe how it behaves.

\Cref{fig:offline_tman}~(a-d) plots the same metrics as seen in \Cref{fig:offline}. The goal is to test how well \sys can improve tuning for workloads with short but many epochs per trial. Here we can observe that \sys also achieves the expected results in this more challenging scenario and reduces both training and tuning time when compared to the baseline systems. Regarding model accuracy, we can also see that our approach achieves comparable or better results than the baseline. Finally, the energy results reflects the performance gains resulting in a more energy efficient approach as well.

To summarize, for these single-tenancy scenarios, \sys presents better performance with up to 23\% reduction on tuning time, is more energy efficient reducing up to 29\% the overall energy consumption of the utilized cluster, and does not affect model accuracy as the observed differences in this aspect are negligible.

\noindent\textbf{Profiling overhead.} Profiling is a fundamental part of our system design and essential for the decision making process. 
During the profiling of a given epoch, the extra computation introduce additional load, depending on the system configuration. 
However, as this profiling overhead only occurs in the epoch granularity and does not apply for all the epochs, the performance benefits resulting from tuning the system-parameters overtake the measured overhead. 
The experimental results presented above also support these assumptions as, otherwise, we would not observe performance gains when compared with the approaches \baselinesys~V1 and \baselinesys~V2 which do not perform any profiling.

\subsection{Multi-Tenancy}

Next, we evaluate \sys in a multi-tenancy scenario (\ie, a shared cluster handling multiple HPT~Jobs). 
In this case, we show the average response time of jobs as an indicator of performance. 
We consider that jobs arrive randomly with the interarrival times being exponentially distributed. 
For the case where two workload types are considered together, each of them corresponds to 50\% of the overall jobs (i.e., equally balanced). 
In all cases, within a given workload type, the workloads are chosen following a round-robin strategy. 
The portion of overall unseen jobs corresponds to 20\%.

\Cref{fig:responsetime} shows the results for the multi-tenancy scenario considering workloads of Type-I and Type-II grouped by type as well as the overall results. 
As in Section~\ref{sec:performance}, this evaluation has been performed in a distributed environment.
In this experiment we observe improvements similar to the ones in the single-tenancy scenario. 
Regarding response time, \sys results in up to 30\% reduction when compared with \baselinesys~V1 and \baselinesys~V2.

\Cref{fig:online_type3} shows the same results described above but considering workloads of Type-III.
This trace was executed in a single node in contrast with the distributed environment of the previously described results.
In this specific scenario we observe that the performance gain trends earlier observed becomes even more evident in such environment and workload type.
In this case, \sys results in up to~65\% reduction on the average response time in comparison with \baselinesys~V1 and \baselinesys~V2.
This indicates that the overhead of computation added for the unseen jobs is compensated by the gain of future similar incoming ones.

%% file: conclusion.tex
%!TEX root = paper.tex
\section{Conclusion}
\label{sec:conclusion}

The combination of hyper and system parameter for Deep Neural Network tuning is an overlooked opportunity that many state-of-the-art tuning solutions ignore.
This paper presented \sys, an open-source system that leverages the repetitive behaviour of DNN tuning jobs to quickly find the best set of parameters. 
Our approach is modular which makes it easy to swap between similarity functions and underlying search algorithms.
We evaluated 7 different real-world datasets from different domains, including text classification and image recognition.
When compared against state-of-the-art DNN tuning systems, \sys shows experimental evidence that the approach greatly reduces tuning and training time while achieving on-par accuracy.